\documentclass{aa}
\usepackage[varg]{txfonts}
\usepackage[hidelinks]{hyperref}
\usepackage{amssymb}
\usepackage{amsmath}
\usepackage{breqn}
\usepackage{subcaption}
\usepackage{natbib}
\usepackage{tabu}
\usepackage{caption}
\usepackage{placeins}
\usepackage{xcolor}

\bibpunct{(}{)}{;}{a}{}{,}

\begin{document}

\title{Spectral characterisation of 14 V-type candidate asteroids from the MOVIS catalogue}

\author{Pavol Matlovič\inst{1}
  \and Julia de Leon\inst{2,}\inst{3}
  \and Hissa Medeiros\inst{2,}\inst{3}
  \and Marcel Popescu\inst{2,}\inst{4}
  \and Juan Luis Rizos\inst{2,}\inst{3}
  \and Jad-Alexandru Mansour\inst{5,}\inst{6}}

\institute{Faculty of Mathematics, Physics and Informatics,
  Comenius University, Bratislava, Slovakia\\
  \email{matlovic@fmph.uniba.sk}  
  \and Instituto de Astrofísica de Canarias (IAC), C/Vía Láctea sn, 38205 La Laguna, Spain
  \and Departamento de Astrofísica, Universidad de La Laguna, 38206 La Laguna, Tenerife, Spain
  \and Astronomical Institute of the Romanian Academy, 5 Cu\c{t}itul de Argint, 040557 Bucharest, Romania
  \and International Centre for Advanced Training and Research in Physics, Magurele 077125, Ilfov, Romania
  \and Faculty of Science and Engineering, University of Groningen, Nijenborgh 9, 9747 AG Groningen}

\date{Received 2020}

\abstract{Most of the currently known basaltic (V-type) asteroids are believed to be past or present members of the Vesta dynamical family. The rising discoveries of V-type asteroids that are not linked to the Vesta family dynamically suggest that a number of major basaltic bodies may have been present during the early stages of the solar system. Using the near-infrared (NIR) colour data in the Moving Objects from VISTA Survey (MOVIS) catalogue, a list of 477 V-type candidates was compiled, with more than half of them outside the Vesta family. In this work, we aim to provide a spectral analysis of 14 V-type candidates of various dynamical types. The computed visible and NIR spectral parameters are used to investigate evidence of space-weathering or mineralogical differences from the expected basaltic composition. Based on the analysis of their visible spectra, we confirm 11 new V-type asteroids: six low-i asteroids - (3188) Jekabsons, (3331) Kvistaberg, (4693) Drummond, (7223) Dolgorukij, (9007) James Bond, and (29733) 1999 BA4; along with four inner-other asteroids - (5524) Lecacheux, (19983) 1990 DW, (51742) 2001 KE$_{55}$, and (90023) 2003 BD$_{13}$; as well as one fugitive - (2275) Cuitlahuac. Additionally, we analysed three peculiar outer main belt V-type candidates based on their visible + NIR spectra. We confirm the diogenite-like composition of (2452) Lyot. The spectrum of asteroid (7302) is not consistent with a basaltic composition and likely reflects an S-type body. The spectrum of (14390) 1990 QP$_{10}$ is similar to a V-type but it shows unique spectral features that suggest a peculiar composition. Overall, our results demonstrate the efficiency of the MOVIS catalogue in identifying V-type objects, with a success rate of over 85\%. The identification of V-types in the inner main-belt is more likely due to the presence of the Vesta family and other nearby asteroids that had escaped from the family. In the middle and outer main belt, where the amount of data is more limited, the proportion of false positives increases.}

\maketitle

%

\section{Introduction}  \label{Introduction}

V-type asteroids are basaltic bodies that are rich in pyroxene and are believed to represent the crust of larger differentiated bodies. Their taxonomic classification is derived from asteroid (4) Vesta, which is the largest 'small body' (530 km in diameter) to show a basaltic crust \citep{1970Sci...168.1445M}. The visible and near-infrared (NIR) spectra of V-type asteroids exhibit two characteristic absorption bands positioned near 1 and 2 $\mu$m. Due to the spectroscopic similarities with basaltic achondrites, Vesta is also considered the parent body of howardites, eucrites, and diogenites, also known as HEDs \citep{2001M&PS...36..501D}. Eucrites are composed of Ca-pyroxene and Ca-plagioclase, the fundamental minerals of basalts. They are one of the extremes in a series that has diogenites (Ca-poor pyroxenes) on the other side. In the middle of this series are the howardites, considered to be breccias composed of debris from both eucrites and diogenites \citep{2004mete.book.....H}.

Most of the discovered V-type asteroids belong to the Vesta collisional family in the inner main belt. Two craters in the south pole of Vesta identified by the Dawn mission \citep{2012Sci...336..690M} most probably represent the origin of the Vesta dynamical family. Since the first discovery of asteroids with spectra similar to Vesta near the 3:1 mean motion resonance with Jupiter \citep{1993Sci...260..186B}, numerous other asteroids with V-type spectra were identified beyond the nominal borders of the family. Based on the dynamical models, these past members of the Vesta family were injected into resonances and moved to different orbits \citep{1991Icar...89....1C, 1994Natur.371..314F}. Today, we know of basaltic asteroids all over the main belt \citep[e.g.][]{2008Icar..193...85N, 2008Icar..198...77M, 2011MNRAS.412.2318D, 2016MNRAS.455.2871I, 2018AJ....156...11H, 2017A&A...600A.126L, 2019MNRAS.488.3866M, 2020MNRAS.491.5966M}. Some of them can be traced back to Vesta through various dynamical pathways \citep{2008Icar..193...85N,2005A&A...441..819C, 2014MNRAS.439.3168C, 2017MNRAS.468.1236B}.

Since the discovery of asteroid (1459) Magnya located in the outer main belt \citep{2000Sci...288.2033L} and with an interpreted bulk composition different from Vesta \citep{2004Icar..167..170H}, the presence of at least another large differentiated object is directly implied. Nowadays, several other basaltic asteroids are known in the middle and outer main belt \citep[][and references therein]{2016MNRAS.455.2871I, 2019MNRAS.488.3866M}. The analysis of oxygen isotopes is particularly useful in meteorites experiencing melting processes at large scales, such as the HEDs. This group typically presents very similar O-isotopic values, which provides a basis for defining an average value and assessing whether deviating values from this average come from having a different parent body as that of the group \citep{2020GeCoA.277..377G}. In this regard, additional evidence arguing in favour of the formation of another HED-like differentiated body in the belt is offered in the discovery of at least two eucrites having distinct, non-HED oxygen isotopic compositions \citep{2002Sci...296..334Y, 2002aste.book..653B, 2017GeCoA.208..145B}.

In this work, we analyse the visible spectra of V-type candidate asteroids selected by \citet{2016A&A...591A.115P} and \citet{2017A&A...600A.126L}, which are presented in Section 2, along with a confirmation of their classification and characteristic spectral properties in Section \ref{VisibleRES}. We further focus on the compositional study of three peculiar asteroids from the outer main belt, for which visible and NIR spectra are analysed in Section \ref{NIRres}. Finally, our conclusions are presented in Section \ref{Conclusions}.

\section{Observations and data reduction}  \label{Observations}

\subsection{Target selection from the MOVIS catalogue}

For the selection of targets for our study, we used the list of V-type candidates identified by \citet{2017A&A...600A.126L}, \citet{2018A&A...617A..12P}, and \citet{2020MNRAS.491.5966M} in the Moving Objects from VISTA survey \citep[MOVIS,][]{2016A&A...591A.115P}. This is a set of three catalogues that include near-infrared colours of 39 947 objects recovered from the Visible and Infrared Survey Telescope for Astronomy \citep[VISTA,][]{2012A&A...548A.119C} and, in particular, its Hemisphere Survey (VISTA-VHS). The VISTA-VHS survey includes three different programs that observe the sky in four filters: Y (1.02 $\mu$m), J (1.25 $\mu$m), H (1.65 $\mu$m), and Ks (2.15 $\mu$m).

It was previously found that the colour-colour plots generated using MOVIS data included clusters that can be associated with different taxonomic classes \citep{2016A&A...591A.115P}. In particular, the (Y-J) versus (J-Ks) colour-colour plot appears to be very efficient in the identification of V-type candidates. The first spectral analyses of V-type candidates from the MOVIS catalogue \citep{2017A&A...600A.126L, 2019MNRAS.488.3866M} have shown high success rates (87 and 83 \%, respectively) for confirmation of the candidate V-types. Approximately half of the proposed V-type candidates selected by \citet{2017A&A...600A.126L} are not members of the Vesta dynamical family.

In the scope of this work, we acquired visible spectra of 14 asteroids from the list of V-type candidates, in particular, those that are not members of the Vesta collisional family. This sample includes 11 asteroids with no previous spectral data and three interesting outer main-belt asteroids with previously analysed NIR spectra. The asteroids studied in this work originate from various dynamical groups, categorised according to the definitions from \citet{2008Icar..193...85N} and \citet{2016MNRAS.455.2871I}. The majority of our sample is found in the inner main belt: six low-i asteroids ($i < 6^{\circ}$, 2.3 $< a < $ 2.5 au), one fugitive ($a <$ 2.3 au and comparable $e$ and $i$ with the Vesta family), and four inner-other asteroids (IOs - asteroids from the inner main belt, which are not members of the Vesta family, low-i or fugitives). The studied sample also includes three outer main-belt bodies categorised as middle and outer main-belt asteroids (MOVs, $a > $ 2.5 au). The orbital and physical parameters of the analysed asteroids are given in Table \ref{properties}.

\begin{table}[]
\centering
\small\begin{center}
\caption{Orbital elements and physical properties of the asteroids analysed in this work: semi-major axis $a$, eccentricity ($e$), inclination ($i$), dynamical group (following \citet{2015aste.book..297N}), and -- when available -- visible albedo ($p_V$) retrieved from the WISE/NEOWISE database \citep{2011ApJ...741...68M, 2012ApJ...759L...8M, 2014ApJ...791..121M, 2016AJ....152...63N}. The orbital and physical parameters were extracted from the JPL Small-Body Database\protect\footnotemark.}
\begin{tabular}{cccccc}
\hline\hline\\[-3mm]
Asteroid & $a$ (au) & $e$ & $i$ ($^{\circ}$) & Group & $p_V$ \\
\hline\\[-3mm]
2275     & 2.296 & 0.170 & 6.393  & Fugitive & 0.177 $\pm$ 0.035   \vspace{0.05cm}\\
2452     & 3.164 & 0.113 & 11.769 & MOV     & 0.380 $\pm$ 0.063   \vspace{0.05cm}\\
3188     & 2.290 & 0.134 & 4.694  & Low-i    & 0.425 $\pm$ 0.071   \vspace{0.05cm}\\
3331     & 2.419 & 0.087 & 3.566  & Low-i    &            -                     \vspace{0.05cm}\\
4693     & 2.279 & 0.083 & 4.866  & Low-i    & 0.366 $\pm$ 0.070   \vspace{0.05cm}\\
5524     & 2.366 & 0.027 & 7.488  & IOs      & 0.034 $\pm$ 0.102   \vspace{0.05cm}\\
7223     & 2.351 & 0.175 & 1.557  & Low-i    & 0.391 $\pm$ 0.073   \vspace{0.05cm}\\
7302     & 2.807 & 0.181 & 9.959  & MOV     & 0.298 $\pm$ 0.026    \vspace{0.05cm}\\
9007     & 2.474 & 0.153 & 5.868  & Low-i    & 0.329 $\pm$ 0.043   \vspace{0.05cm}\\
14390    & 3.244 & 0.103 & 6.560  & MOV    & 0.220 $\pm$ 0.057    \vspace{0.05cm}\\
19983    & 2.390 & 0.058 & 6.379  & IOs       & 0.335 $\pm$ 0.044   \vspace{0.05cm}\\
29733    & 2.447 & 0.087 & 5.542  & Low-i    & 0.214 $\pm$ 0.119   \vspace{0.05cm}\\
51742    & 2.340 & 0.143 & 8.078  & IOs       &                 -                 \vspace{0.05cm}\\
90223    & 2.459 & 0.114 & 7.149  & IOs       &                 -                 \vspace{0.05cm}\\
\hline
\end{tabular}
\label{properties}
\end{center}
\end{table}

\subsection{Data acquisition and reduction}

The  visible spectra presented here were mostly obtained using the Intermediate Dispersion Spectrograph (IDS) instrument mounted on the 2.54m Isaac Newton Telescope (INT), as part of standard programs C67/2018B and  C94/2019A as well as the Alhambra Faint Object Spectrograph and Camera (ALFOSC) at the 2.5m Nordic Optical Telescope (NOT), under the service program SST2016-346, both located at the El Roque de Los Muchachos Observatory (ORM) in La Palma (Canary Islands, Spain). A summary of the observational circumstances for each studied asteroid is given in Table \ref{conditions}. 

\begin{table*}[]
\centering
\small\begin{center}
\caption{Observational circumstances of the asteroids analysed in this work: telescope used for the observations, date and starting time of the first exposure (UT start), total exposure time (T$_{exp}$), airmass, apparent magnitude ($m_{V}$), phase angle ($\alpha$), and heliocentric distance at the time of the observation ($r_{h}$). The last column includes information on the solar analogue stars used to obtain the asteroid reflectance spectra.}
\begin{tabular}{ccccccccl}
\hline\hline\\[-3mm]
Asteroid & Telescope & Date and UT start & T$_{exp}$ (s) & Airmass  & $m_{V}$ & $\alpha$ ($^{\circ}$) & $r_{h}$ (au) & Solar analogues \\
\hline\\[-3mm]
2275      & INT       & 20/11/18 \, 04:35 & 3x300   & 1.346  & 15.0 & 6.5    & 2.120 & SA93-101, Hyades64, SA98-978 \vspace{0.05cm}\\
2452      & NOT     & 08/12/18 \, 04:50 & 3x300   & 1.436  & 16.3 & 4.8    & 3.306 & SA98-978, SA102-1081 \vspace{0.05cm}\\
3188      & INT       & 23/04/19 \, 23:21 & 3x1200 & 1.904  & 16.8 & 5.0    & 2.504  & SA107-998, SA110-361 \vspace{0.05cm}\\
3331      & NOT     & 08/12/18 \, 05:28 & 3x600   & 1.112   & 17.7 & 26.2  & 2.222 & SA98-978, SA102-1081 \vspace{0.05cm}\\
4693      & INT       & 23/04/19 \, 21:18 & 3x1200 & 1.141  &  17.2 & 20.2  & 2.258  & SA107-998, SA110-361\vspace{0.05cm}\\
5524      & NOT     & 08/12/18 \, 06:13 & 3x600   & 1.254   & 17.6 & 24.7  & 2.357 & SA98-978, SA102-1081 \vspace{0.05cm}\\
7223      & INT       & 27/08/18 \, 05:02 & 2x900   & 1.023   & 17.0 & 27.9  & 1.957  & SA112-1333, SA115-271, HD5495 \vspace{0.05cm}\\
              &              &                            &              &              &         &         &            & HD9761, HD10861, HD167221, HD204433 \vspace{0.05cm}\\
7302      & INT       & 27/08/19 \, 00:56 & 3x600   & 1.483   & 15.7 & 3.4   & 2.567  & SA112-1333, SA115-271 \vspace{0.05cm}\\
9007      & INT       & 27/08/18 \, 02:40 & 3x900   & 1.108   & 17.2 & 21.8  & 2.117  & SA112-1333, SA115-271, HD5495\vspace{0.05cm}\\
              &              &                            &              &              &         &         &            & HD9761, HD10861, HD167221, HD204433 \vspace{0.05cm}\\
14390    & INT       & 27/08/19 \, 00:17 & 3x600   & 1.524   &16.4  & 4.4   & 2.915 & SA112-1333, SA115-271 \vspace{0.05cm}\\
19983    & INT       & 20/11/18 \, 05:37 & 4x500   & 1.195   &16.5  & 13.5  & 2.251 & SA93-101, Hyades64, SA98-978 \vspace{0.05cm}\\
29733    & INT       & 23/05/19 \, 01:54 & 3x1200 & 1.510   &16.4 & 2.4   & 2.265  & SA107-998, SA112-1333, HD89165 \vspace{0.05cm}\\
51742    & INT       & 23/05/19 \, 01:03 & 3x1200  & 1.305  &17.2 & 4.9   & 2.030  & SA107-998, SA112-1333, HD89165 \vspace{0.05cm}\\
90223    & INT       & 23/05/19 \, 03:08 & 2x1800 & 1.468   &17.6 & 6.1 & 2.182 & SA107-998, SA112-1333, HD89165 \vspace{0.05cm}\\
\hline
\end{tabular}
\label{conditions}
\end{center}
\end{table*}

For the acquisition of the visible spectra with the IDS spectrograph, we used the RED+2 CCD detector with the low-resolution R150V grating, which covers the 0.40--0.95 $\mu$m spectral interval. A 1.2\arcsec slit was employed, oriented along the parallactic angle to correct for differential refraction effects. The tracking was set at the asteroid's proper motion, extracted from the corresponding ephemeris. Observations with the ALFOSC spectrograph were performed using the CCD\#14, a 2048x2064 back illuminated detector providing a 0.214\arcsec/pixel scale, and the Grism\#4 (0.32--0.91 $\mu$m). A second-order blocking filter (GG420) was used to avoid contamination at red wavelengths. The final set-up covered a wavelength range from 0.42 to 0.91 $\mu$m. A 2.5\arcsec slit was used, oriented in the parallactic angle, and the tracking was also set at the asteroid's proper motion, as in the case of observations with the IDS.

A total of two to four spectra were obtained for each target and subsequently averaged to increase the quality of the final spectra. Several solar analogue stars were observed each night at airmass similar to those of the targets. Data reduction followed the standard procedures and was done using IRAF tasks \citep{1986SPIE..627..733T}. The spectra of both the target asteroids and solar analogues were bias- and flat-field-corrected, extracted from the sky background and collapsed to one dimension. Then, the spectra were wavelength-calibrated using the CuAr+CuNe lamps (IDS) or ThAr+Ne+He lamps (ALFOSC). Finally, each asteroid spectrum was divided by all available solar analogue spectra and the results were normalised to unity at 0.55 $\mu$m and averaged to obtain a final reflectance spectrum. The resulting visible spectra for the 14 V-type candidates are displayed in Fig. \ref{spectra}.

\footnotetext{https://ssd.jpl.nasa.gov/sbdb.cgi}

\section{Visible spectra analysis}  \label{VisibleRES}

\subsection{Spectral parameters and taxonomy}

To analyse the visible spectra of our target asteroids, we computed three parameters which characterise the position and shape of the 0.9 $\mu$m absorption band: the reflectance gradient in the 0.50 -- 0.75 $\mu$m range (Slope A), the reflectance gradient in the 0.80 -- 0.92 $\mu$m range (Slope B), and the ratio between the reflectance at 0.75 and 0.90 $\mu$m (apparent depth). These parameters were defined in \citet{2016MNRAS.455.2871I} and computed for a sample of V-types in the Vesta collisional family (also called vestoids), V-type near-Earth asteroids (NEAs), and another four dynamical groups that were described in the previous section: fugitives, low-i, IOs, and MOVs. Their statistical results provided characteristic values of these parameter for the different populations.

The errors of Slope A and Slope B were obtained as a difference between the values obtained in the predefined wavelength ranges (0.50 -- 0.75 $\mu$m and 0.80 -- 0.92 $\mu$m respectively) and restricted wavelength ranges of 0.55 -- 0.72 $\mu$m for \textit{Slope A} and 0.80 -- 0.90 $\mu$m for \textit{Slope B}, thus accounting both for the spectra noisiness and subtle wavelength calibration errors. The error of the \textit{apparent depth} was obtained as a standard deviation of five independent manual measurements.

\begin{figure*}[t]
\begin{center}
\includegraphics[width=0.95\columnwidth,angle=0]{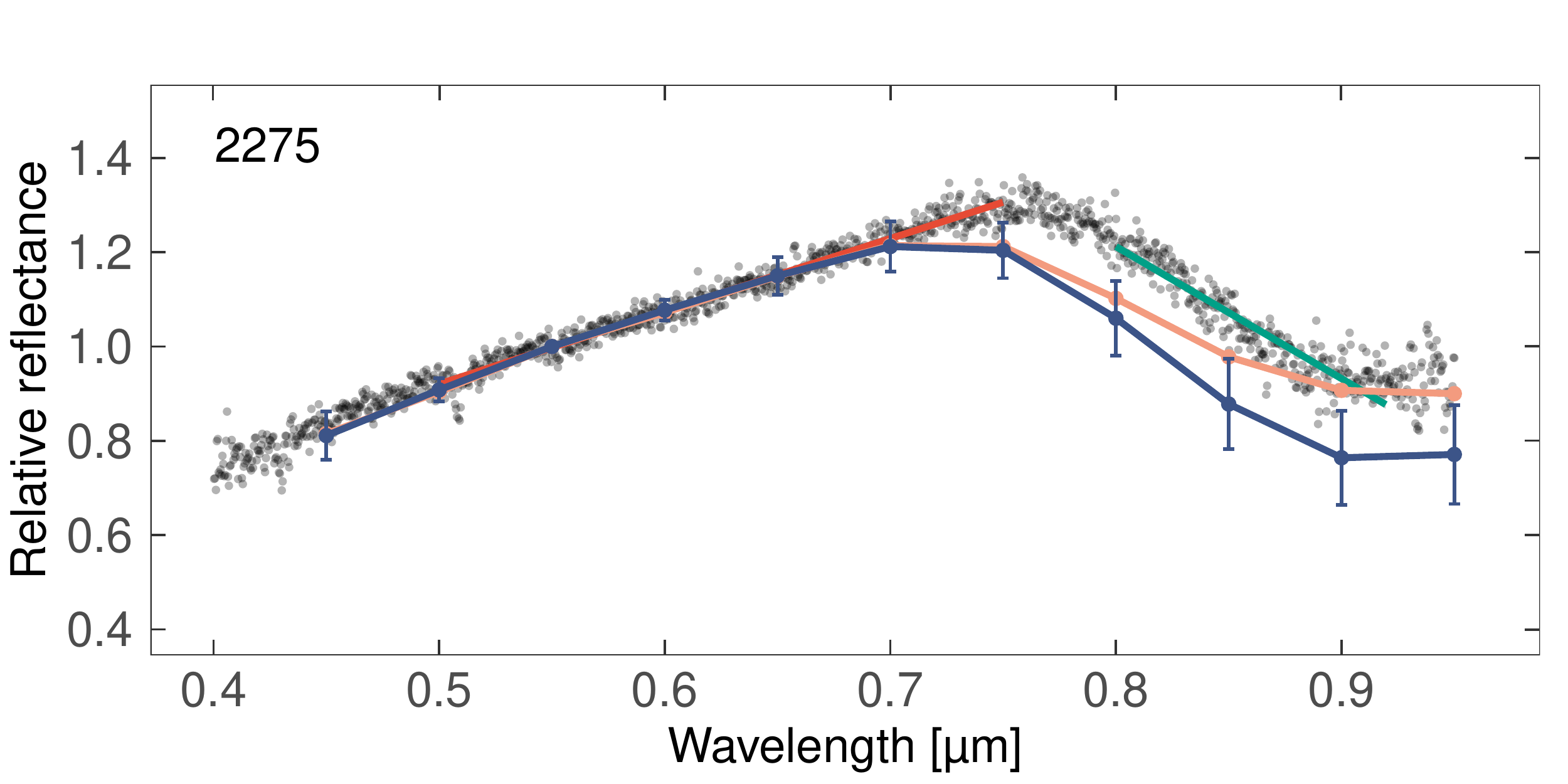}
\includegraphics[width=0.95\columnwidth,angle=0]{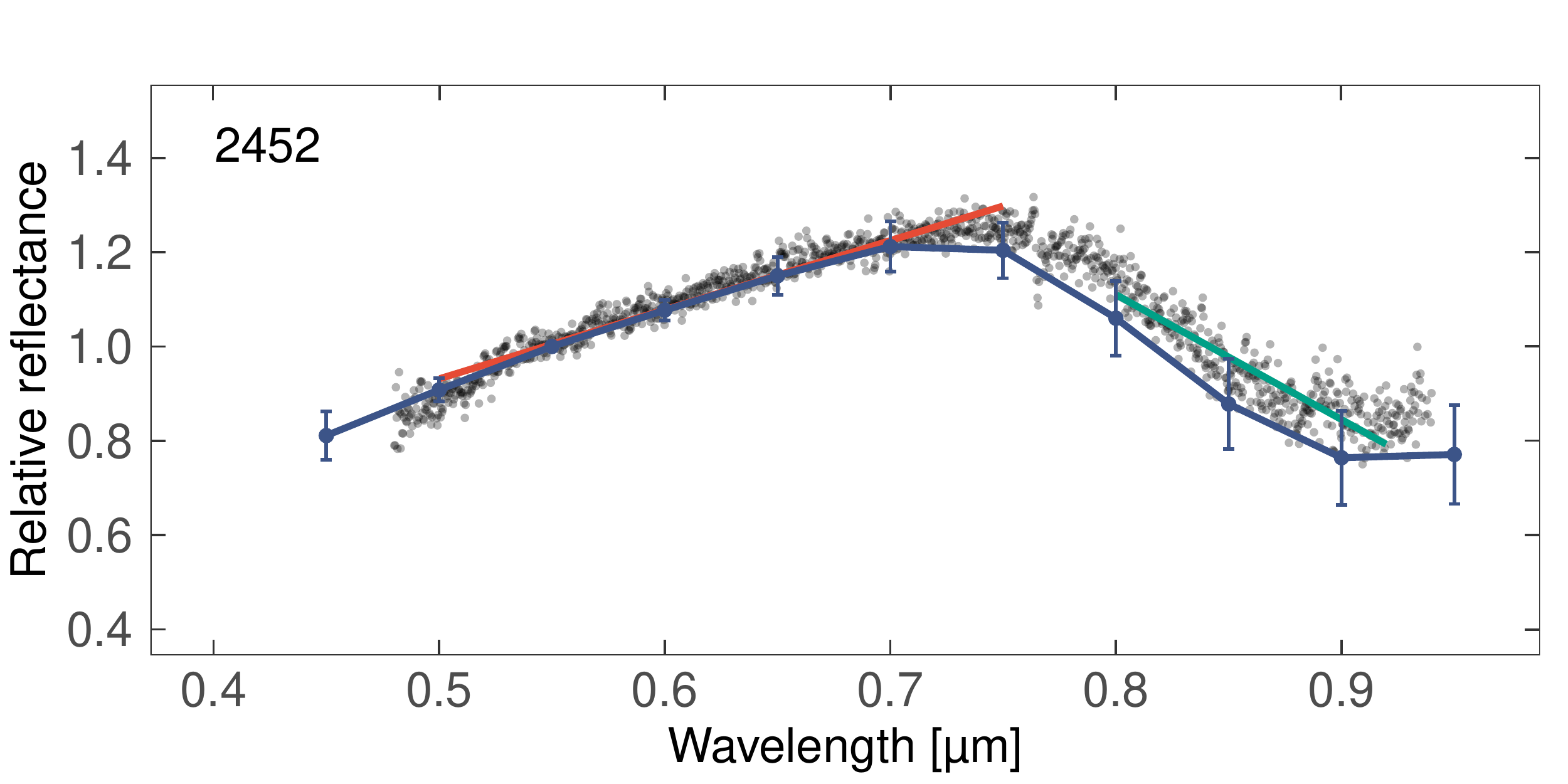}
\includegraphics[width=0.95\columnwidth,angle=0]{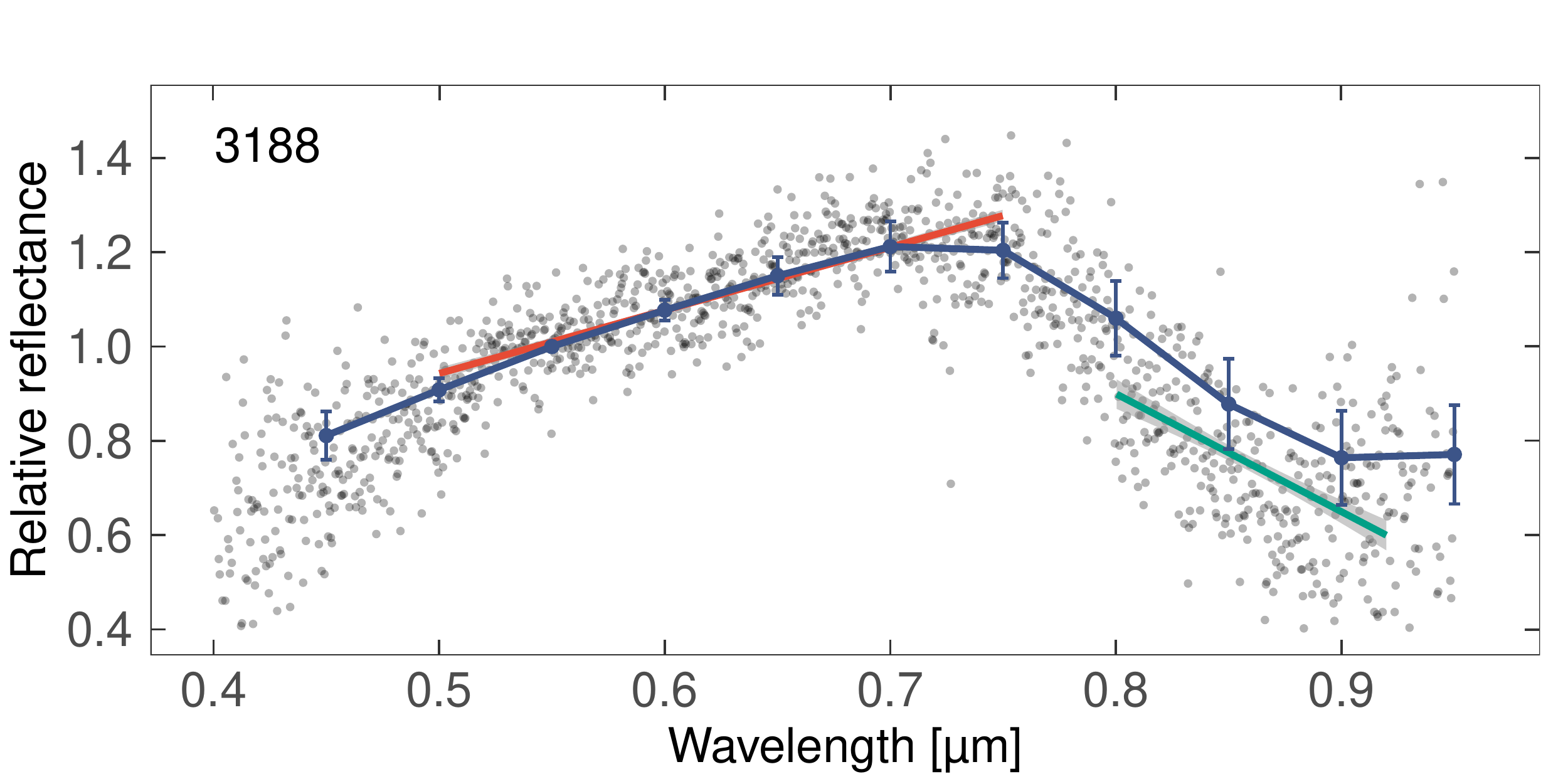}
\includegraphics[width=0.95\columnwidth,angle=0]{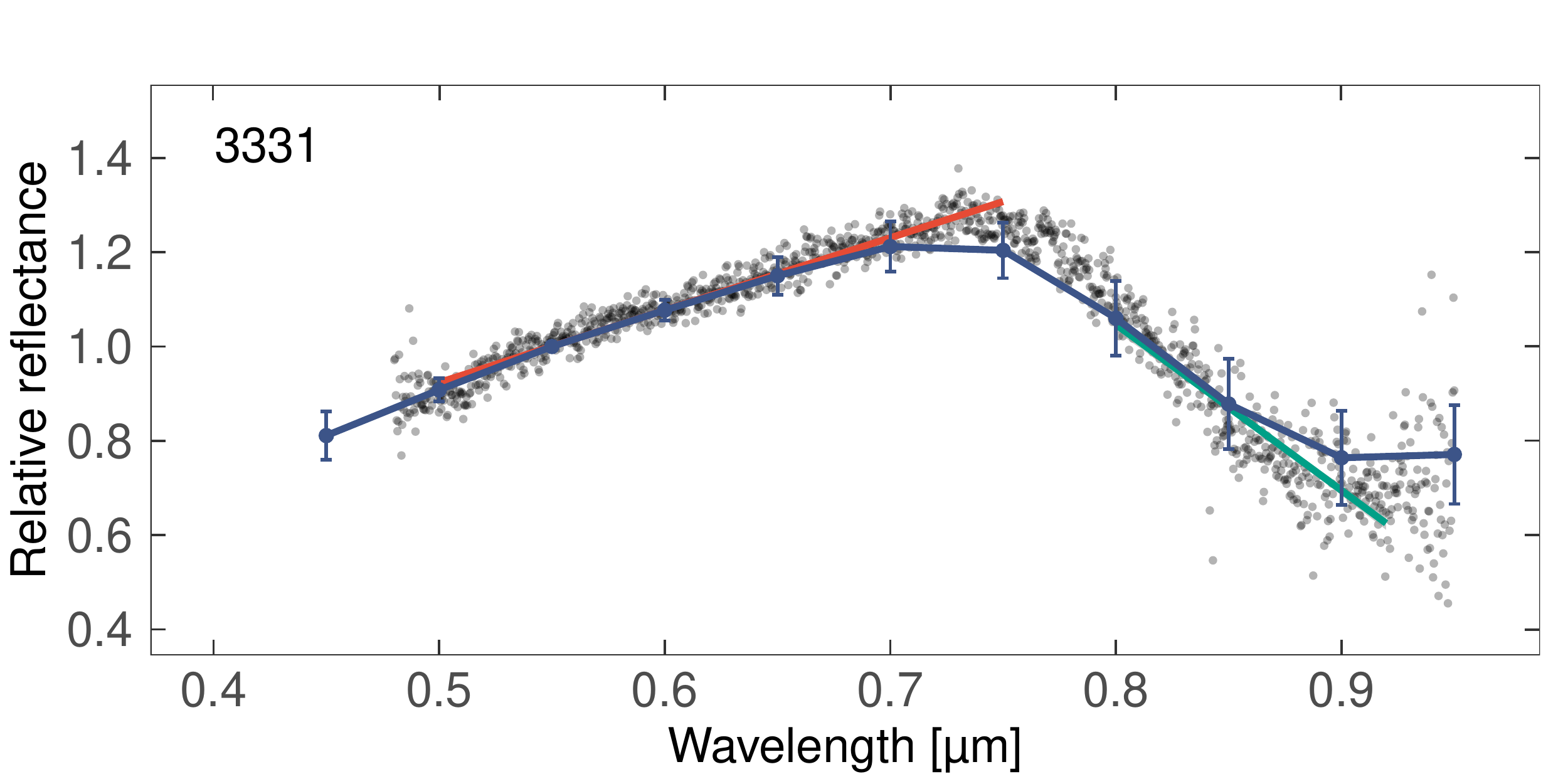}
\includegraphics[width=0.95\columnwidth,angle=0]{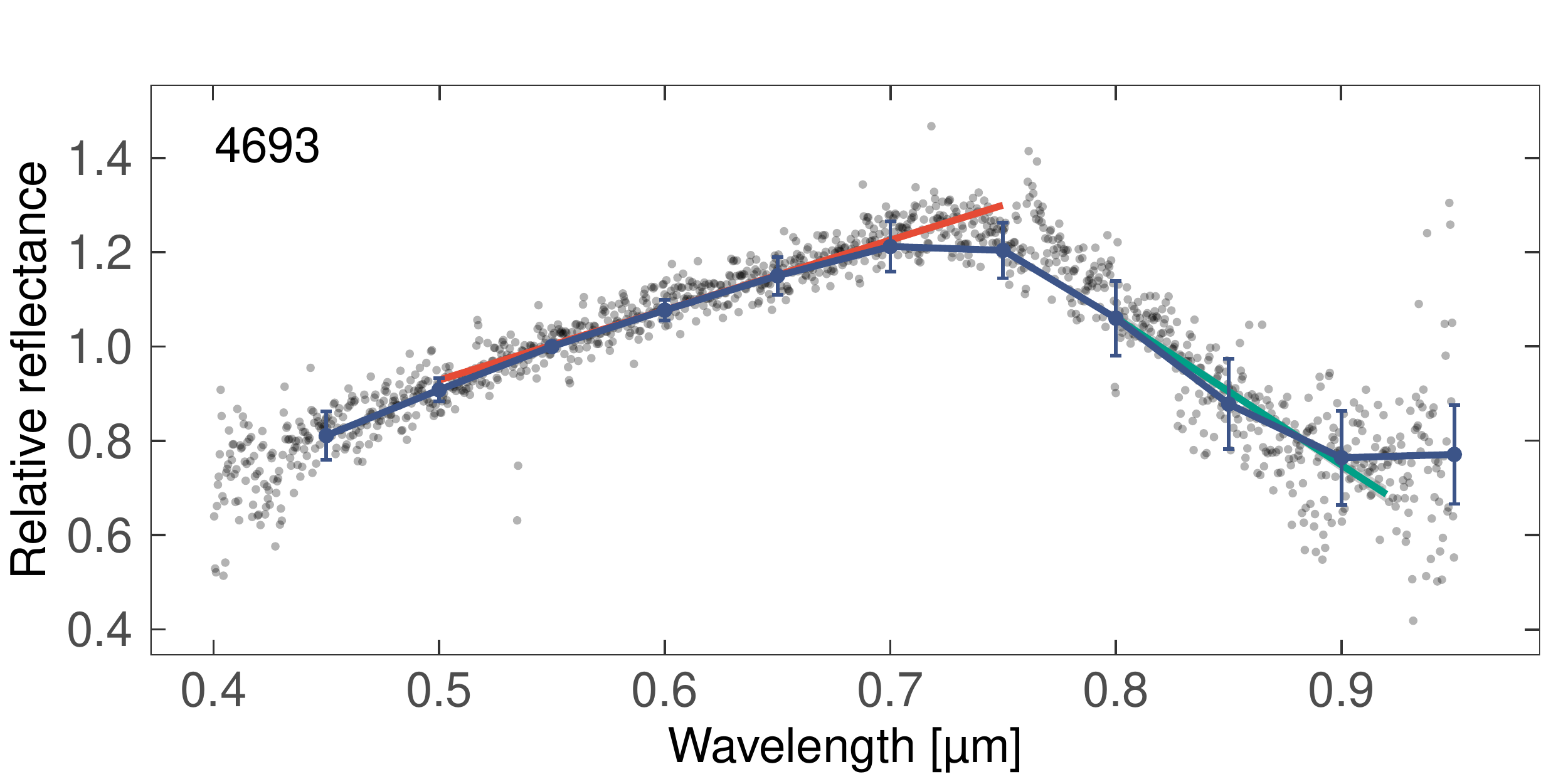}
\includegraphics[width=0.95\columnwidth,angle=0]{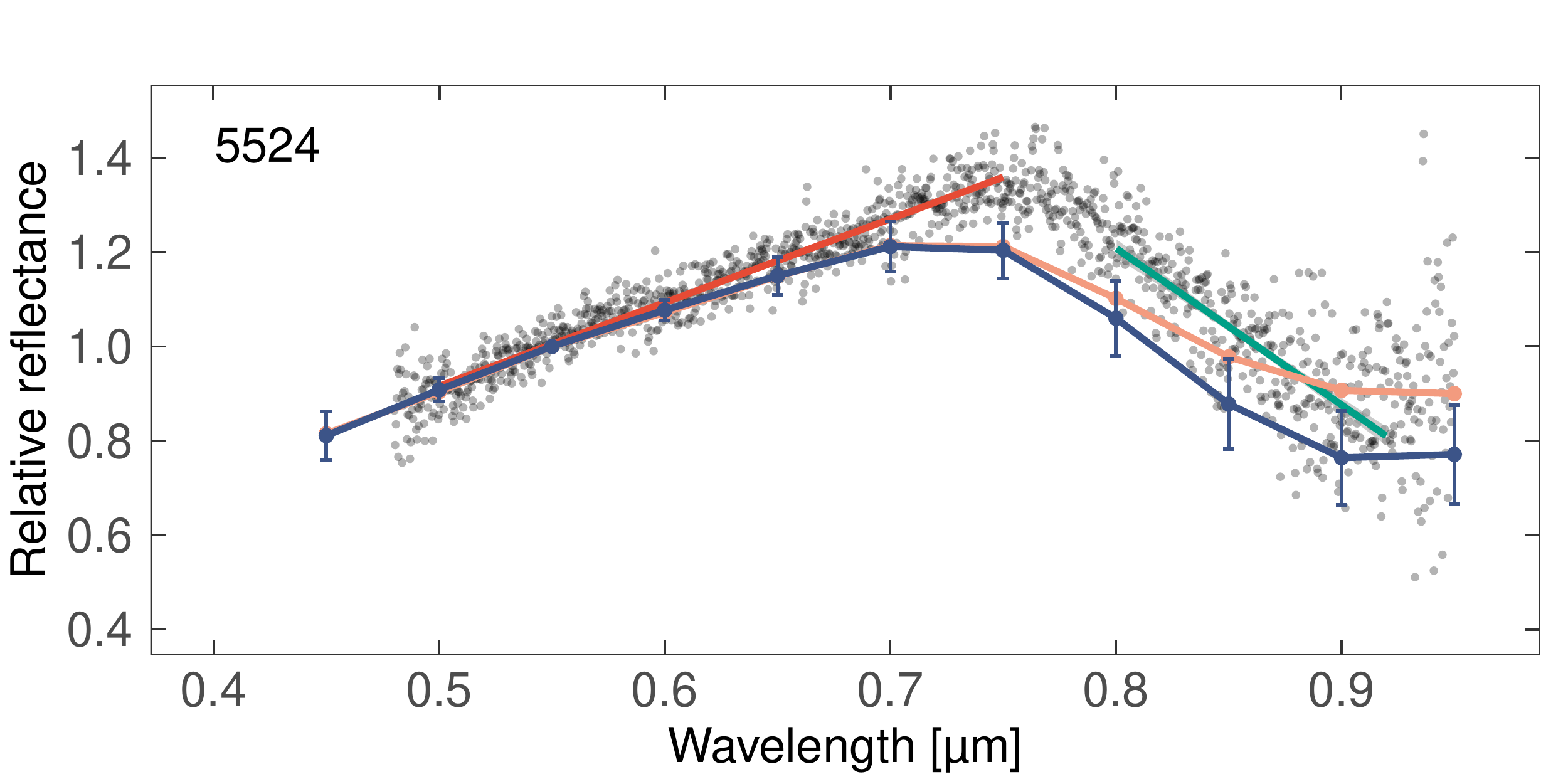}
\includegraphics[width=0.95\columnwidth,angle=0]{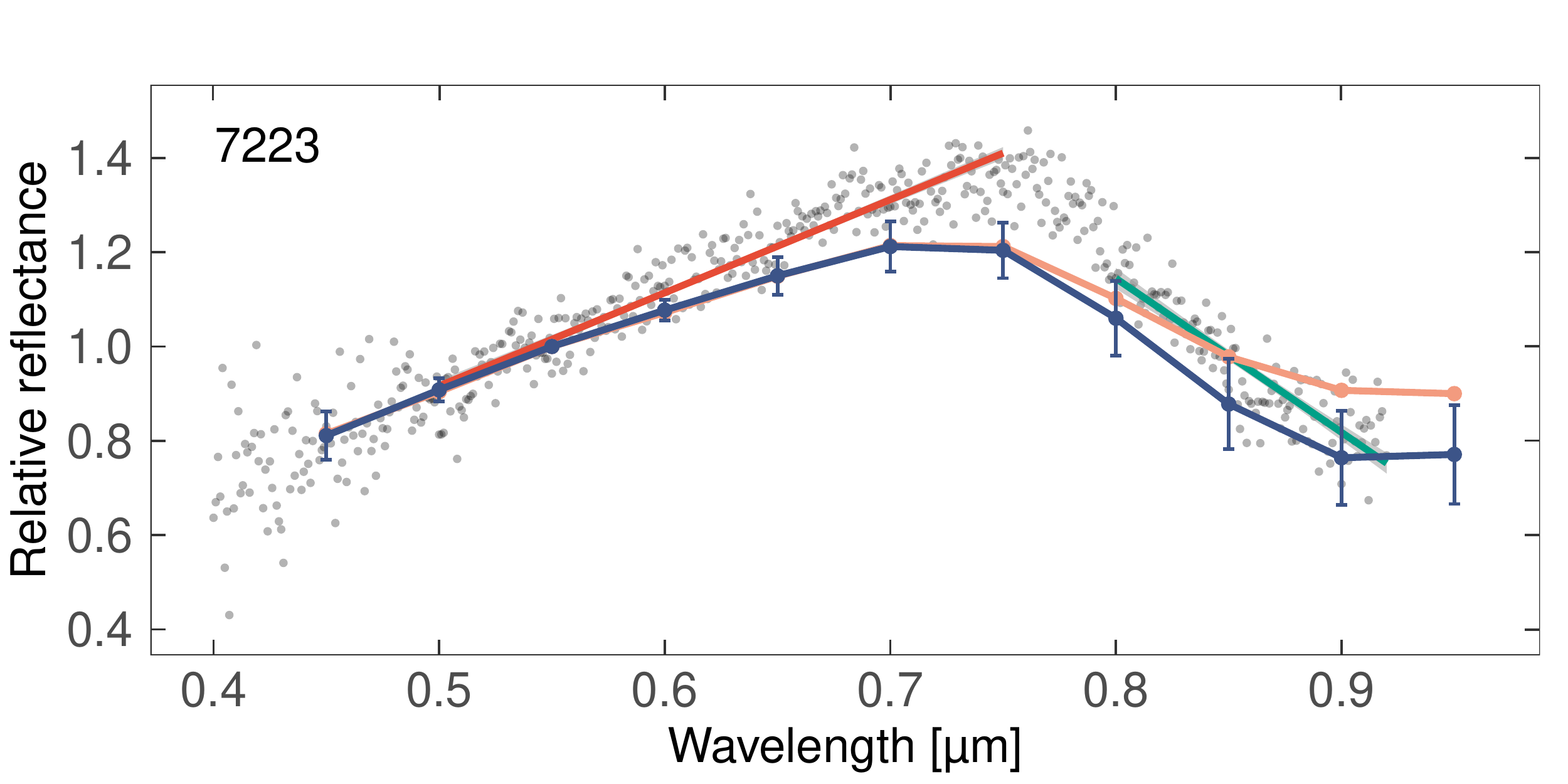}
\includegraphics[width=0.95\columnwidth,angle=0]{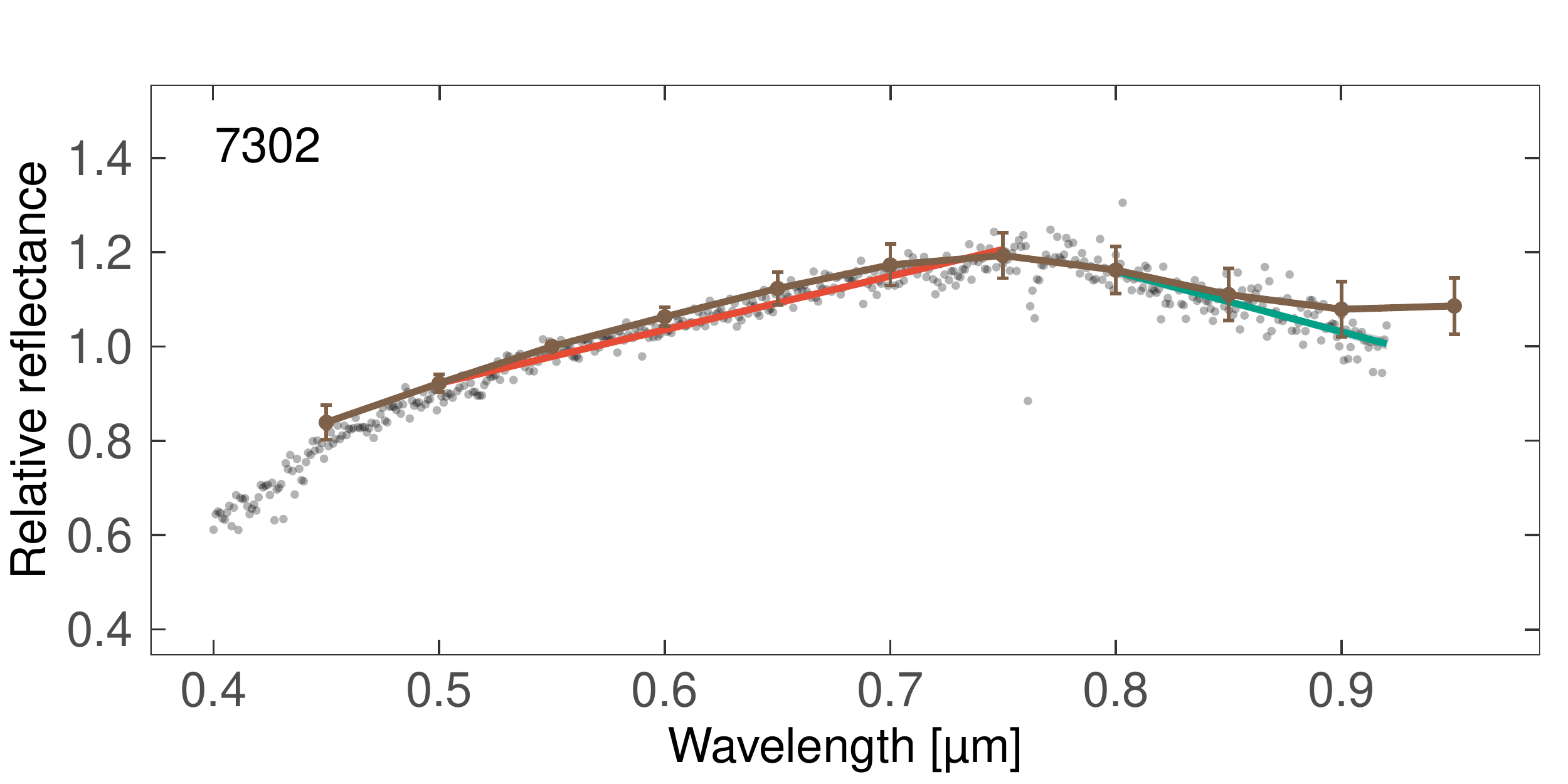}
\end{center}
\caption{Visible spectra of studied V-type candidate asteroids (grey points). The red and green lines represent reflectance gradient fits used for the computation of the \textit{Slope A} and \textit{Slope B} parameters. We also plot the average spectrum corresponding to the V-type taxonomic class and its standard deviation (in blue) as defined by \citet{2009Icar..202..160D}. In some cases, average R-type (in orange) or S-type (in brown) spectrum from \citet{2009Icar..202..160D} are also plotted for comparison.}%
\label{spectra}%
\end{figure*}

\begin{figure*}[t]
\ContinuedFloat
\begin{center}
\includegraphics[width=0.95\columnwidth,angle=0]{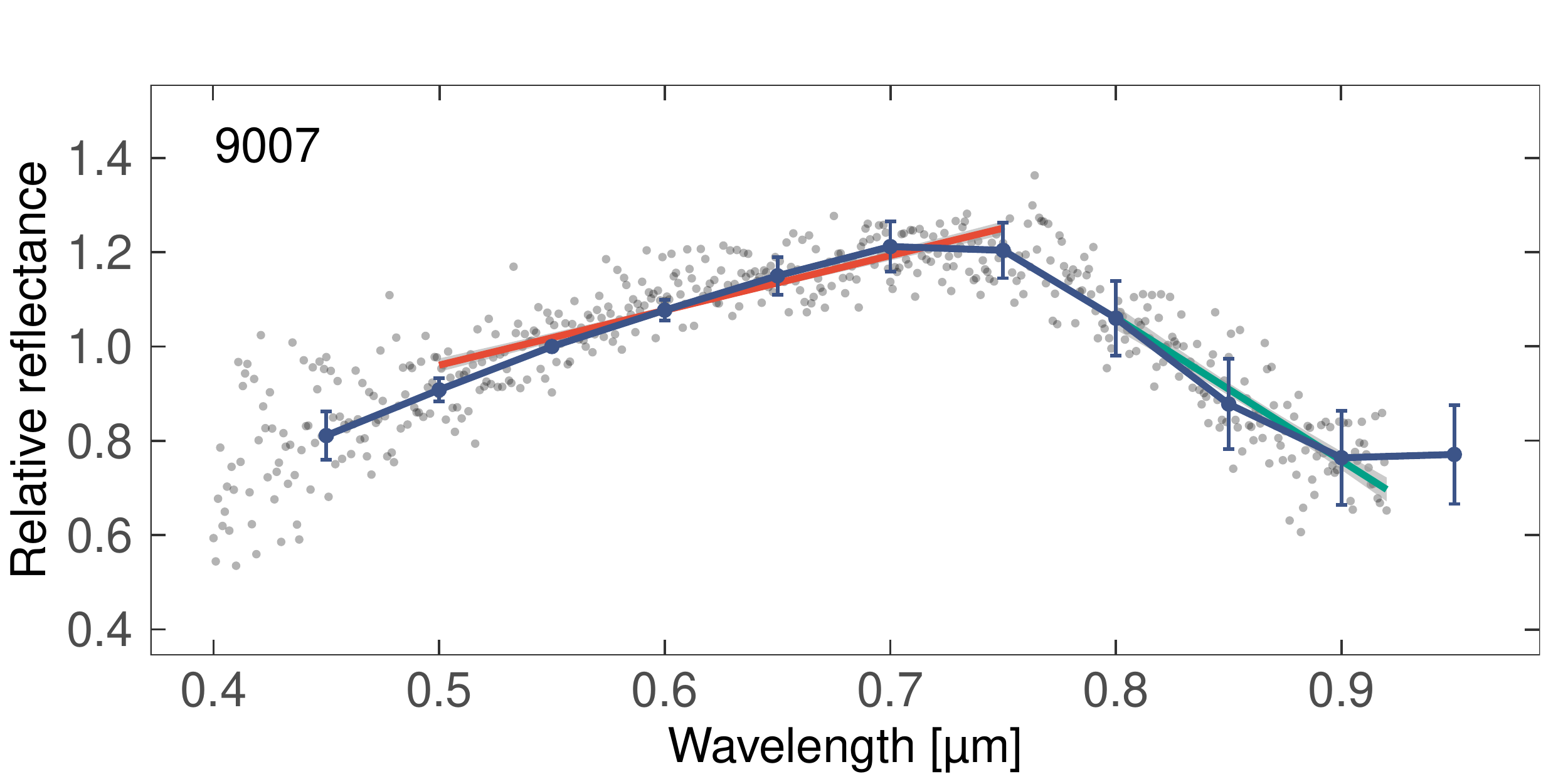}
\includegraphics[width=0.95\columnwidth,angle=0]{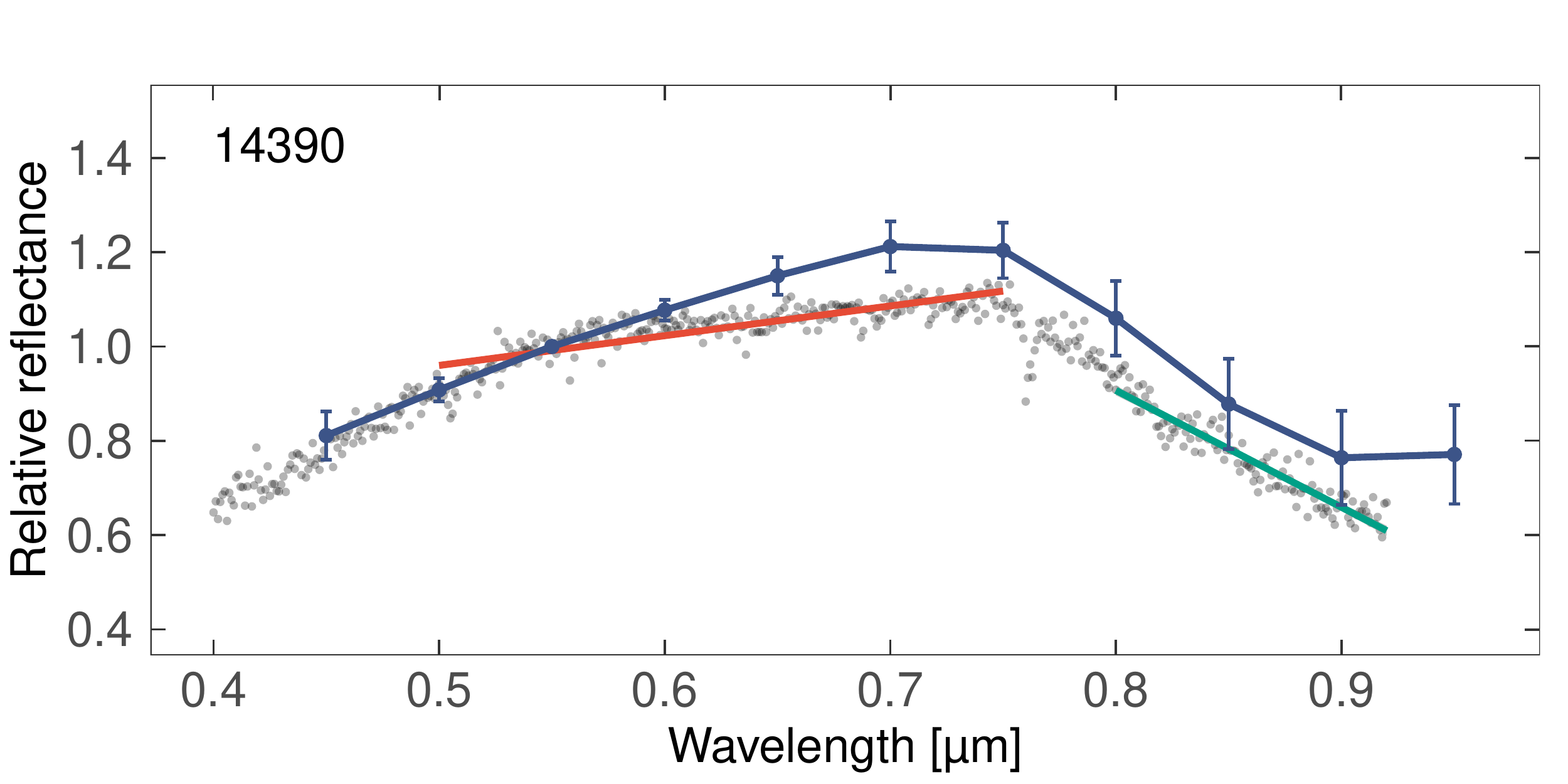}
\includegraphics[width=0.95\columnwidth,angle=0]{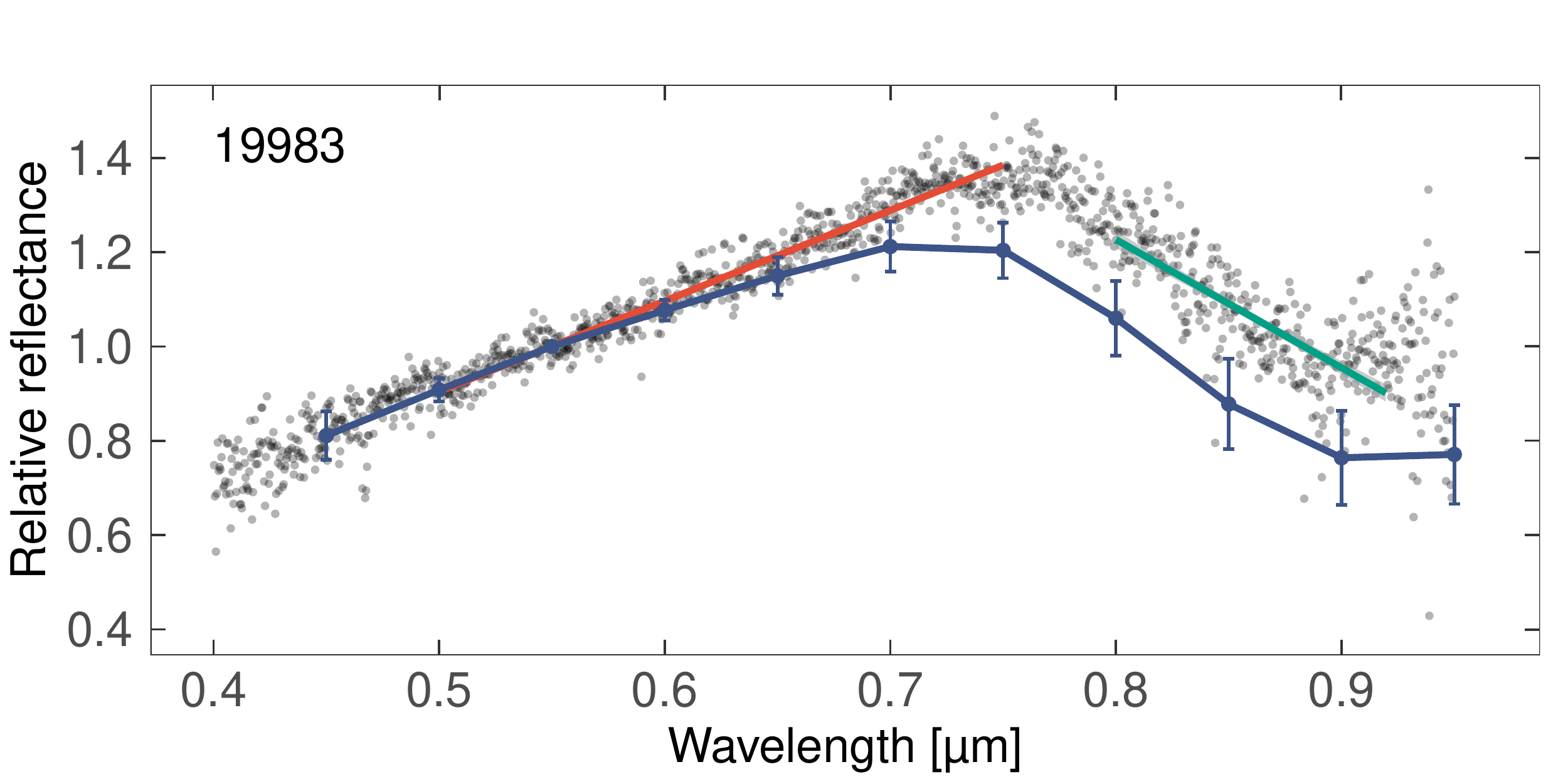}
\includegraphics[width=0.95\columnwidth,angle=0]{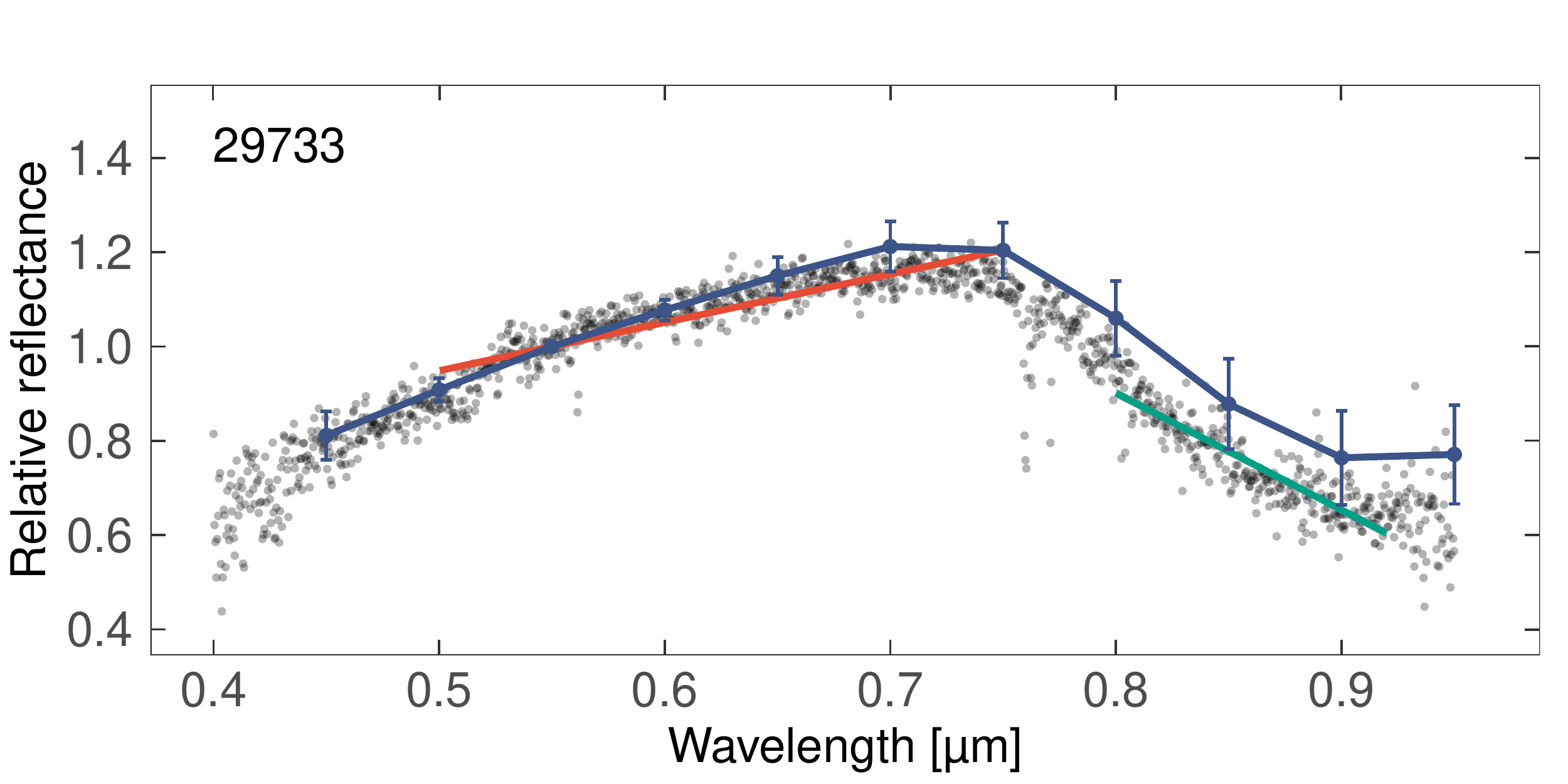}
\includegraphics[width=0.95\columnwidth,angle=0]{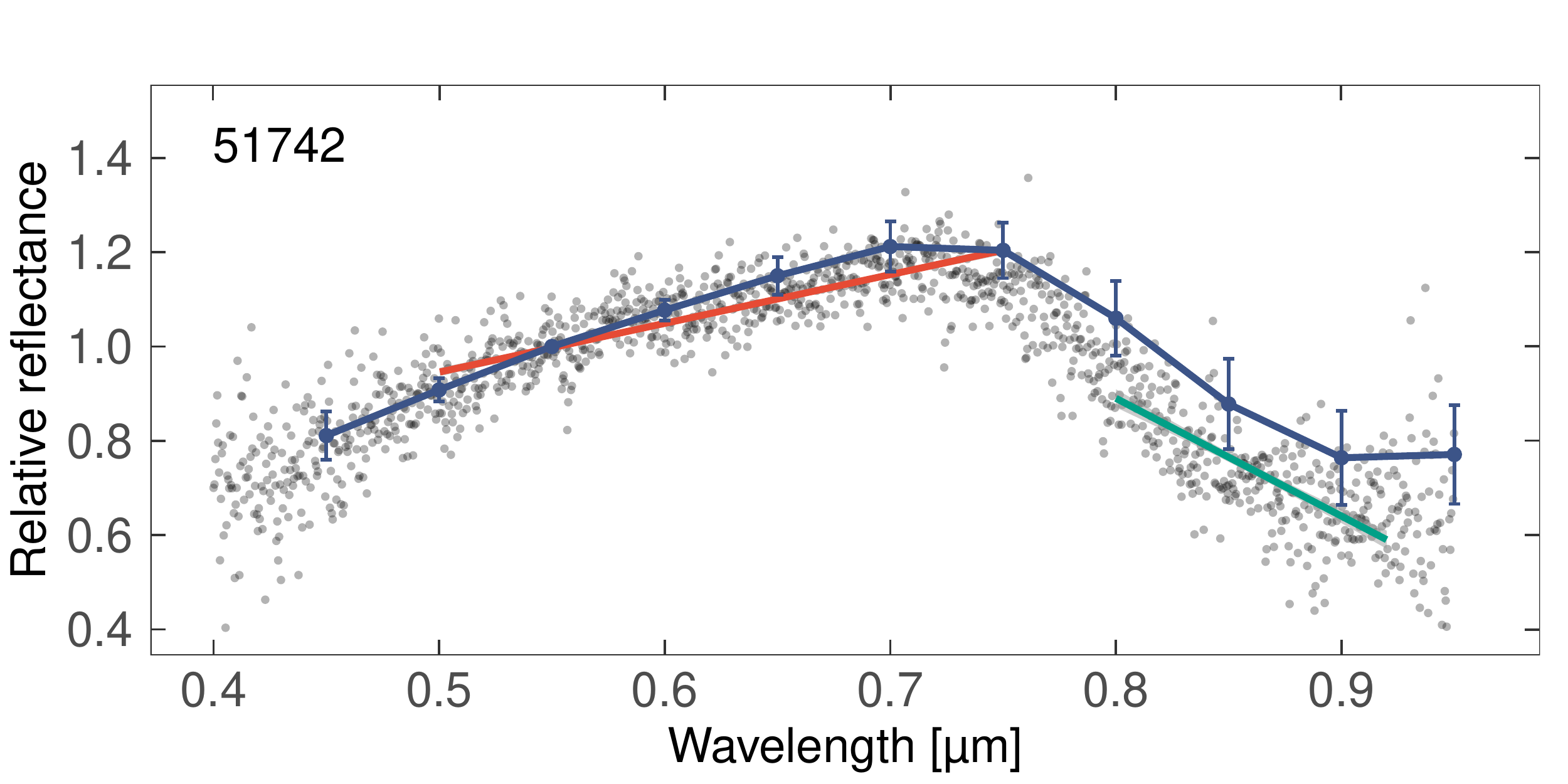}
\includegraphics[width=0.95\columnwidth,angle=0]{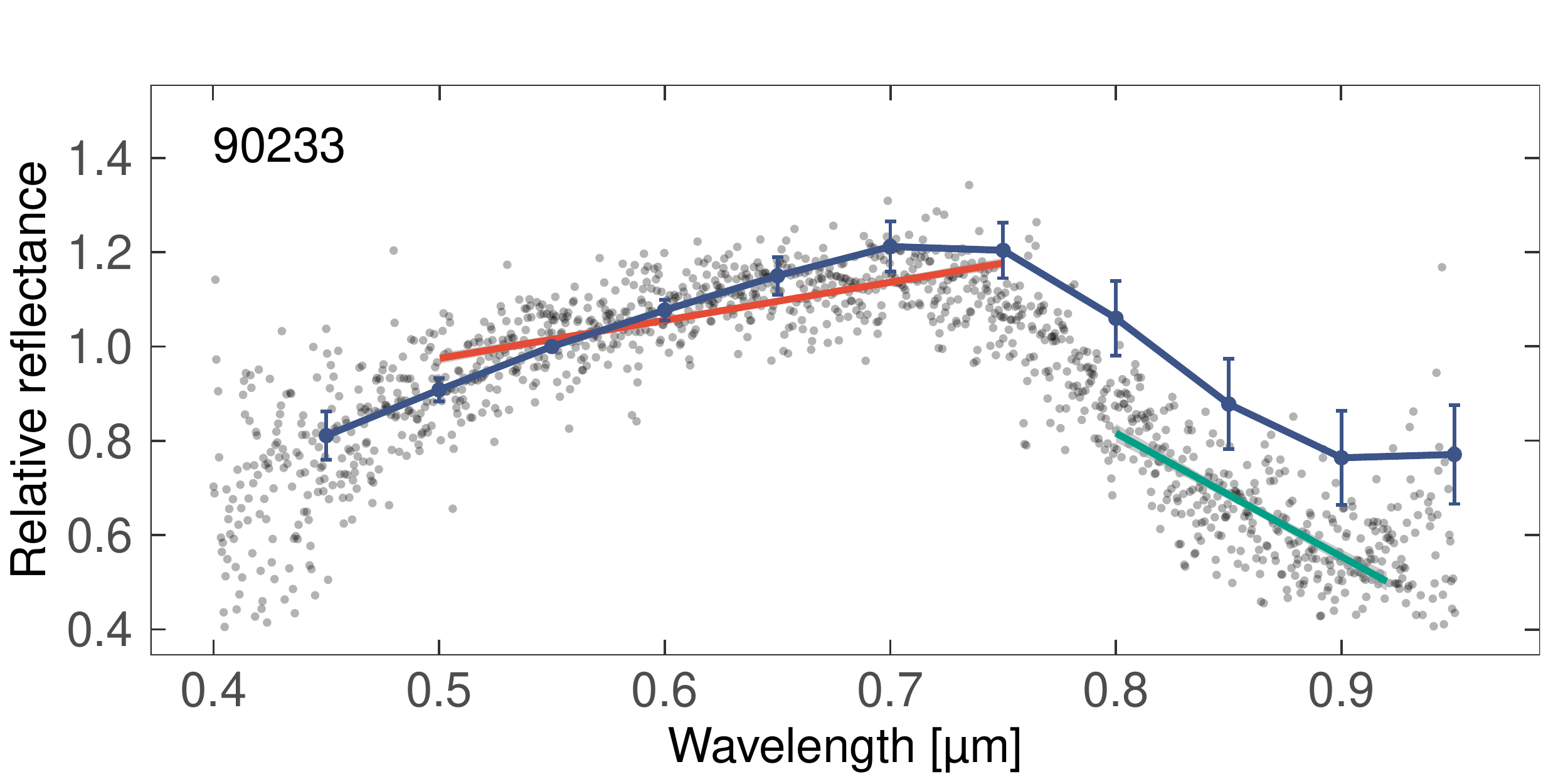}
\end{center}
\caption[]{(continued)}%
\label{spectra}%
\end{figure*}

We also calculated the \textit{Slope A} parameter corrected for the phase angle effect (\textit{Slope A corr}), following the empirical relation by \citet{2012Icar..217..153R}:

\begin{equation}
S_c = S - \gamma \alpha, \quad \alpha$ $\lesssim$ 25$^{\circ}
,\end{equation}
where $S$ is the measured spectral slope, $\alpha$ is the phase angle, and $\gamma$ is the reddening coefficient $\gamma$ = 0.0198 \%/$\mu$m/deg.

The phase angle effect can cause some objects to appear redder as the phase angle increases, affecting the observed reflectance slopes. All of the studied asteroids in this work were observed at phase angles $\alpha \lesssim$ 30$^{\circ}$, compatible with the \citet{2012Icar..217..153R} relation. We note that this relation was inferred for Vesta and we assume that V-types present a similar mineralogy. As we cannot discard that some of the target asteroids might present differences in composition when compared to asteroid 4 Vesta, both values of \textit{Slope A} and \textit{Slope A corr} were considered for the analysis. The determined spectral parameters for each asteroid are given in Table \ref{slopes}.

\begin{table*}[]
\centering
\small\begin{center}
\caption{Visible spectra parameters of the analysed asteroids: \textit{Slope A}, \textit{Slope B}, \textit{Slope A corr} - which is \textit{Slope A} corrected for the phase angle effect, and the apparent depth. The closest taxonomic type is given based on the measured parameters (Tax.), using the M4AST tool \citep{2012A&A...544A.130P} and for asteroids with NIR spectra also using the SMASS tool.}
\begin{tabular}{cccccccc}
\hline\hline\\[-3mm]
Asteroid & \textit{Slope A} & \textit{Slope A corr} & \textit{Slope B} & App. depth & Tax.& M4AST & SMASS \\
               & (\%/0.1$\mu$m) & (\%/0.1$\mu$m)  &  (\%/0.1$\mu$m) &                 &       &              \\
\hline\\[-3mm]
2275  & 15.28 $\pm$ 0.70 & 13.99 $\pm$ 0.70& -29.59 $\pm$ 1.57 & 1.468 $\pm$ 0.026 & V & R or Sa  &   \vspace{0.05cm}\\
2452  & 14.66 $\pm$ 0.46 & 13.71 $\pm$ 0.46& -26.50 $\pm$ 3.64 & 1.532 $\pm$ 0.047 & V & V  & V       \vspace{0.05cm}\\
3188  & 13.40 $\pm$ 1.21 & 12.41 $\pm$ 1.21& -25.02 $\pm$ 9.44 & 1.743 $\pm$ 0.123 & V & V  &         \vspace{0.05cm}\\
3331  & 15.36 $\pm$ 0.67 & 10.17 $\pm$ 0.67& -35.19 $\pm$ 4.57 & 1.841 $\pm$ 0.122 & V & V  &         \vspace{0.05cm}\\
4693  & 14.84 $\pm$ 0.37 & 10.84 $\pm$ 0.37& -31.21 $\pm$ 6.45 & 1.597 $\pm$ 0.160 & V & V  &         \vspace{0.05cm}\\
5524  & 17.73 $\pm$ 0.86 & 12.84 $\pm$ 0.86& -33.24 $\pm$ 5.07 & 1.609 $\pm$ 0.036 & V & R or Sa &    \vspace{0.05cm}\\
7223  & 20.69 $\pm$ 0.43 & 15.17 $\pm$ 0.43& -34.65 $\pm$ 5.57 & 1.664 $\pm$ 0.128 & V & R or V  &    \vspace{0.05cm}\\
7302  & 11.37 $\pm$ 0.24 & 10.70 $\pm$ 0.24& -12.78 $\pm$ 1.78 & 1.157 $\pm$ 0.030 & S & S & S        \vspace{0.05cm}\\
9007  & 12.88 $\pm$ 1.15 & 8.56  $\pm$ 1.15& -33.64 $\pm$ 4.75 & 1.558 $\pm$ 0.155 & V & V &          \vspace{0.05cm}\\
14390 & 6.31  $\pm$ 1.40 & 5.44  $\pm$ 1.40& -24.78 $\pm$ 1.97 & 1.685 $\pm$ 0.043 & ? & V or O & V   \vspace{0.05cm}\\
19983 & 19.21 $\pm$ 2.87 & 16.54 $\pm$ 2.87& -27.20 $\pm$ 1.39 & 1.426 $\pm$ 0.105 & V & R or Sa &    \vspace{0.05cm}\\
29733 & 10.24  $\pm$ 1.07 & 9.76  $\pm$ 1.07& -24.80 $\pm$ 3.27 & 1.761 $\pm$ 0.093 & V & V &         \vspace{0.05cm}\\
51742 & 10.34 $\pm$ 0.52 & 9.37  $\pm$ 0.52& -24.01 $\pm$ 6.07 & 1.710 $\pm$ 0.066 & V & V &          \vspace{0.05cm}\\
90223 & 7.94  $\pm$ 0.72 & 6.73  $\pm$ 0.72& -25.73 $\pm$ 4.89 & 1.985 $\pm$ 0.101 & V & V &          \vspace{0.05cm}\\
\hline
\end{tabular}
\label{slopes}
\end{center}
\end{table*}

A taxonomical classification of the presented asteroids was performed independently based on the measured visible spectra parameters and using the on-line tool M4AST\footnote{http://spectre.imcce.fr/m4ast/index.php} developed for asteroid spectra modelling \citep{2012A&A...544A.130P}. The M4AST tool fits a polynomial curve to the observed asteroid spectrum and finds the closest taxonomical type based on the smallest mean squared error from the average representative spectra from the Bus-DeMeo taxonomy \citep{2009Icar..202..160D}. For those asteroids having NIR spectra (2452, 7302, and 14390), we used both the M4AST tool and the on-line classification tool available at the web page of the Planetary Spectroscopy Group of the MIT\footnote{http://smass.mit.edu/}. The classification procedures are slightly different: while the M4AST uses a curve-matching approach, the MIT tool applies a spline fit to smooth the spectrum, samples the 0.45-2.45 $\mu$m range into 41 data points, computes and removes the continuum (or slope), and, finally, applies a principal component analysis (PCA). The results are shown in the last column of Table \ref{slopes}.

While near-infrared spectra are necessary for a mineralogical characterisation of asteroid surfaces \citep{1986JGR....9111641C, 1993Icar..106..573G, 2002aste.book..183G}, some information can be also inferred from visible spectra parameters. The effects of meteoroid and micrometeoroid impacts, irradiation and sputtering of solar wind particles and cosmic rays, altogether referred to as space weathering, can cause an alteration of the surface mineralogy of asteroids. Basaltic asteroids generally respond to these effects by exhibiting a redder spectral slope, lower albedo, and shallower spectral bands. Therefore, the detection of a steeper \textit{Slope A} can indicate surfaces affected by space weathering \citep{2012A&A...537L..11F,2016MNRAS.455..584F}, while greater apparent depth could reflect a larger grain size, the presence of unweathered pyroxene, or a different mineralogy \citep{2013Icar..223..850C, 2017ApJS..233...14F}.

\subsection{Results and discussion}

\begin{figure*}[t]
\begin{center}
\includegraphics[width=.45\textwidth]{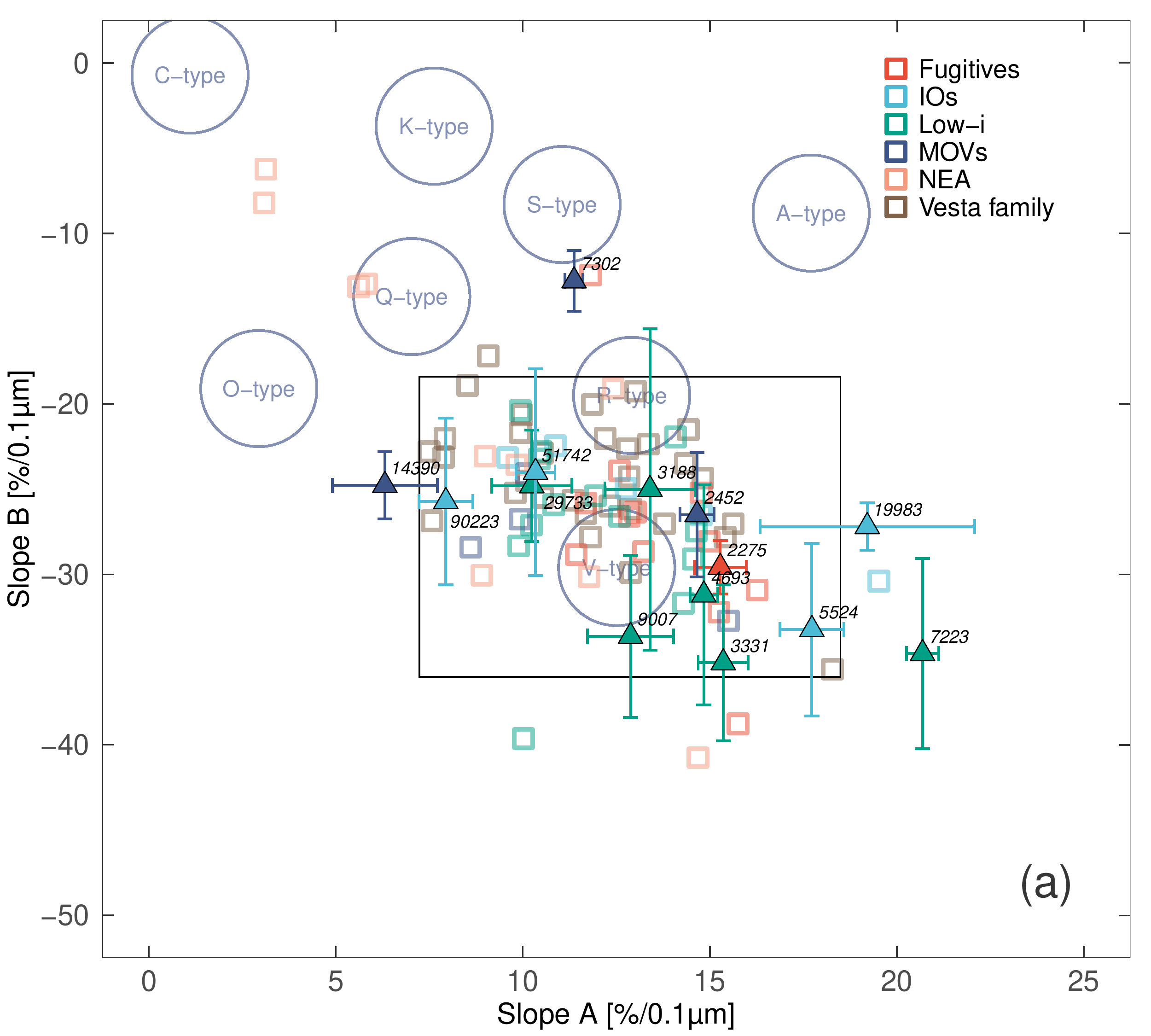}
\includegraphics[width=.45\textwidth]{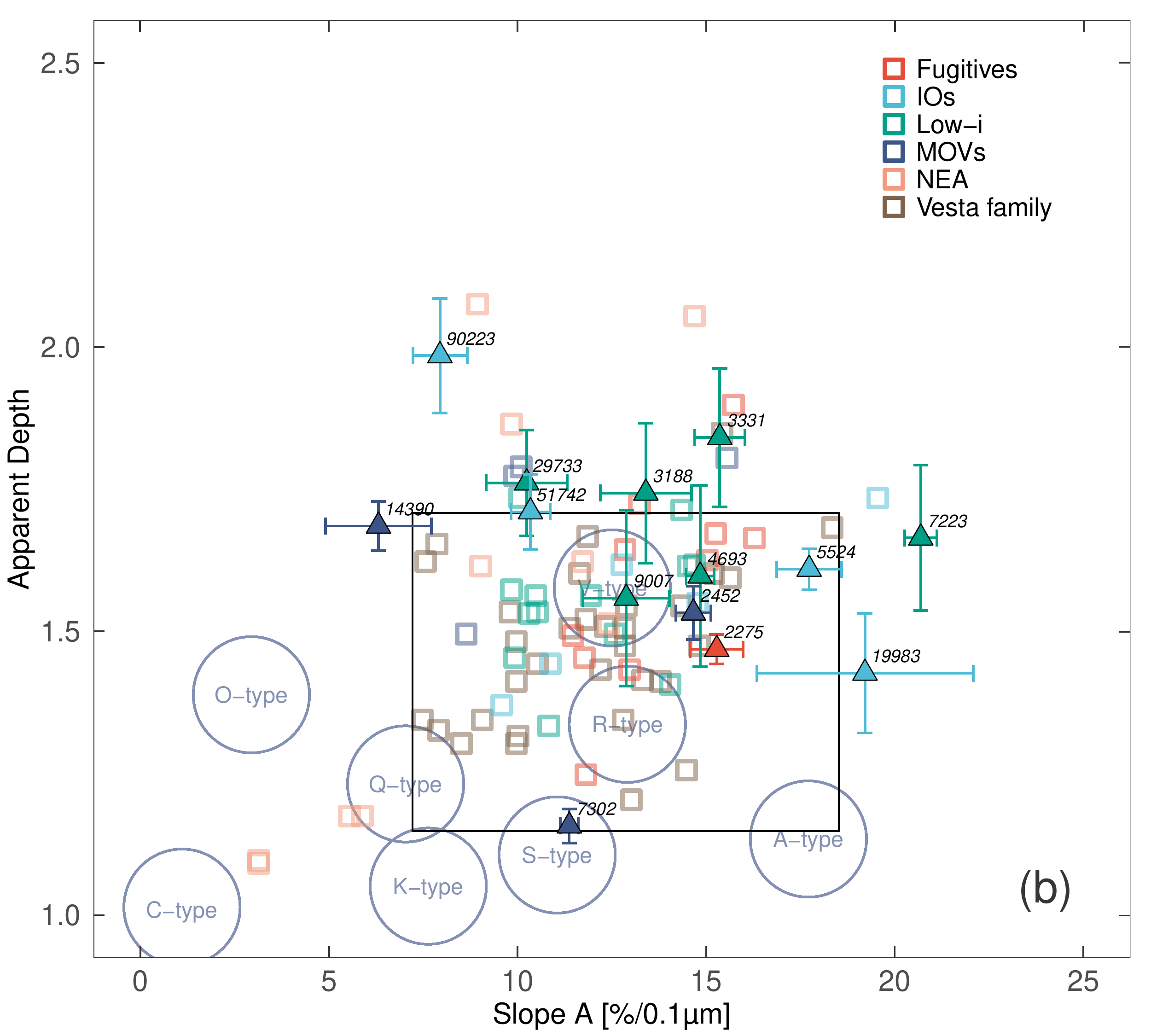}
\includegraphics[width=.45\textwidth]{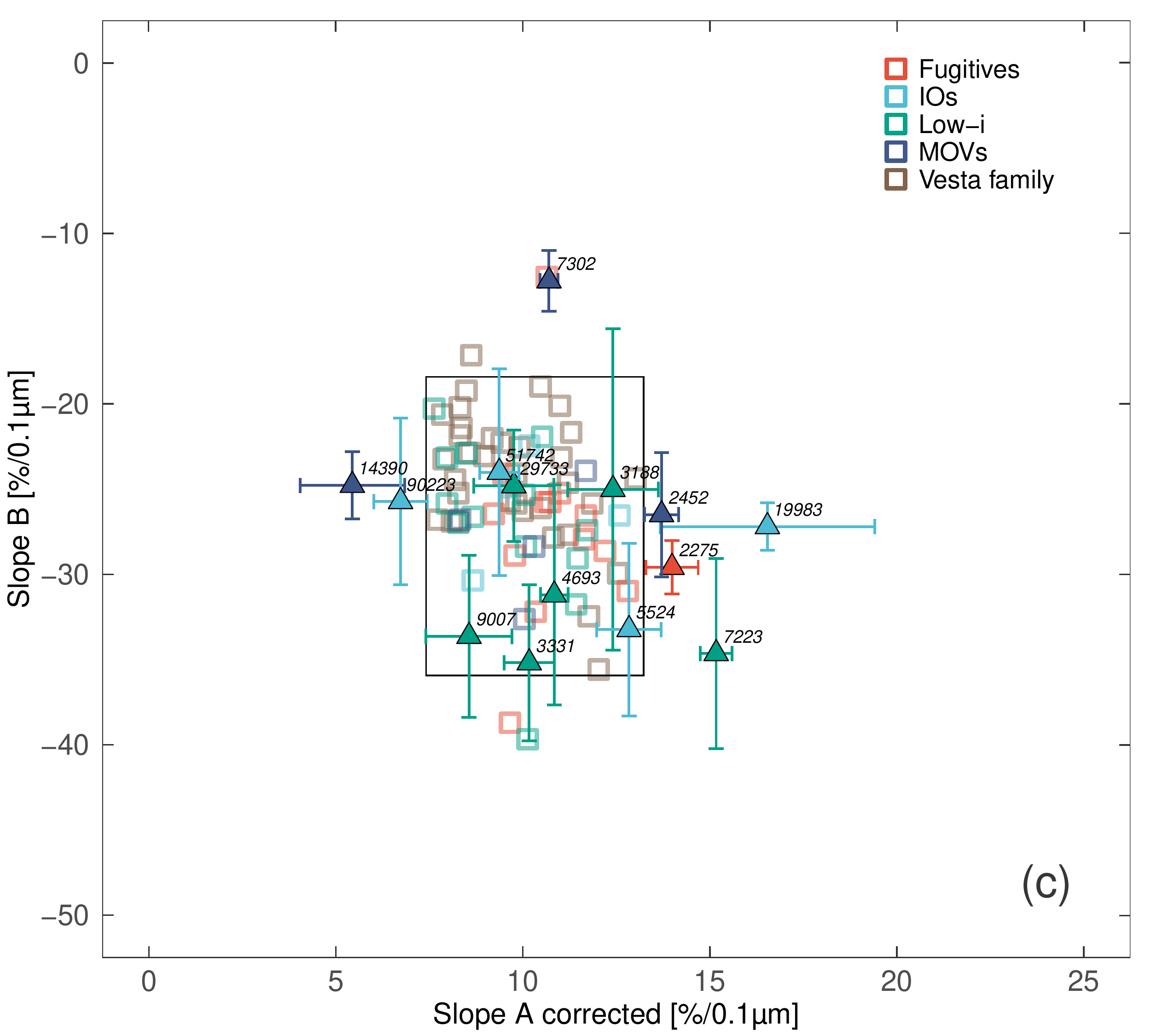}
\includegraphics[width=.45\textwidth]{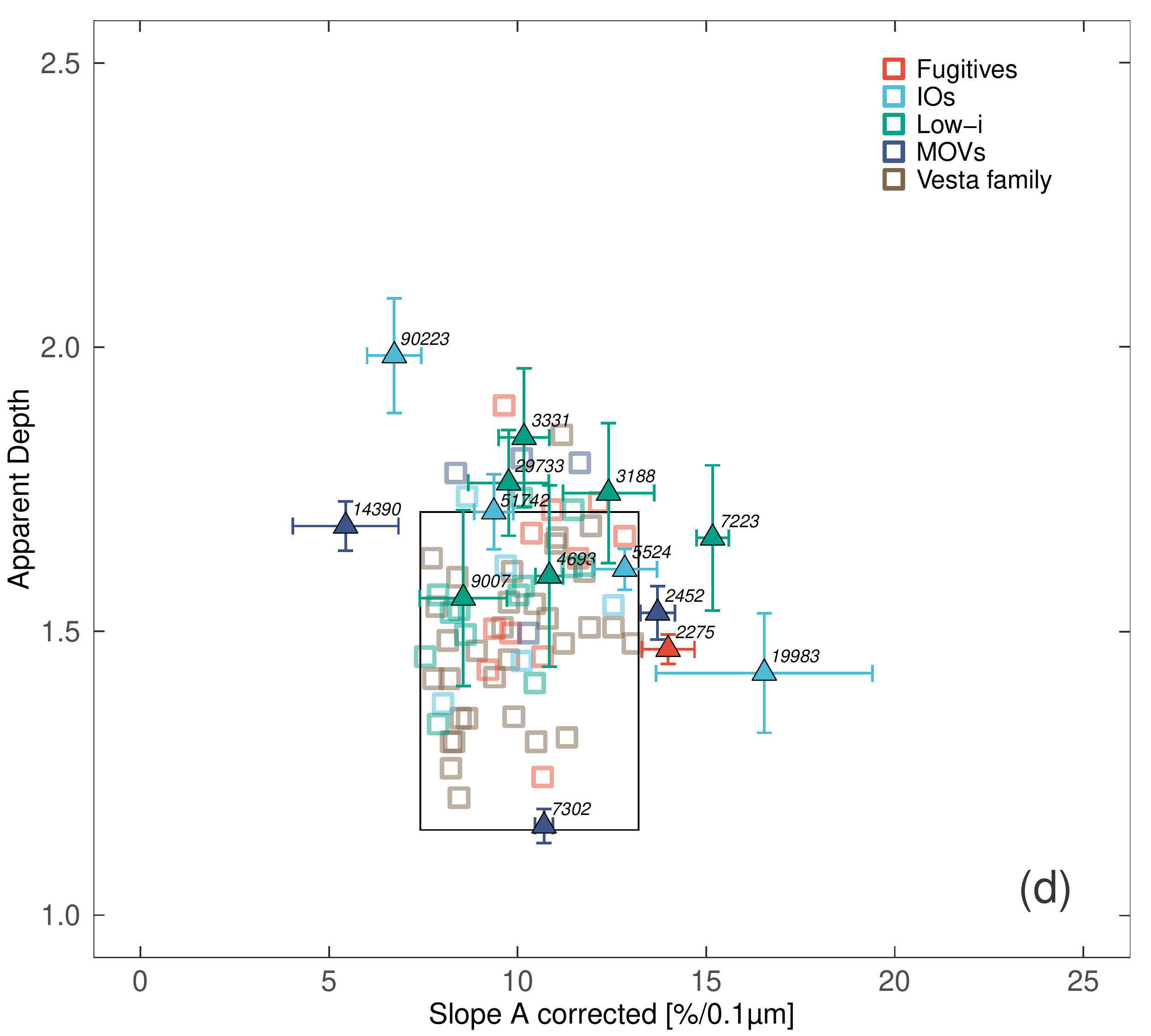}
\end{center}
\caption{\textit{Slope A} versus \textit{Slope B} (a) and apparent depth (b) of the analysed V-type candidate asteroids (triangles). The same parameters are represented versus \textit{Slope A corr} in (c) and (d), respectively. The studied samples are plotted along with the results from the statistical analysis of V-type asteroids by \citet{2016MNRAS.455.2871I} (squares). The box region delimits parameters of a V-type control sample from \citet{2016MNRAS.455.2871I}. Colour coding differentiates between the dynamical groups of asteroids. Circles in the top panels represent average positions of other taxonomic types based on the mean spectra from the Bus-DeMeo catalogue \citep{2009Icar..202..160D}, with diameter of the circle estimated from the \textit{Slope A} error based on the one-sigma of the mean S-type spectrum.}%
\label{SlopesDepths}%
\end{figure*}

Figure \ref{SlopesDepths} displays the measured slopes and apparent depths of the studied asteroids and their comparison to V-type asteroids as analysed by \citet{2016MNRAS.455.2871I}. The box regions displayed in Fig. \ref{slopes} delimit the spectral parameters for V-type control sample from \citet{2016MNRAS.455.2871I}: 7.26 \%/0.1$\mu$m < \textit{Slope A} < 18.51 \%/0.1$\mu$m, -18.41 \%/0.1$\mu$m < \textit{Slope B} < -35.98 \%/0.1$\mu$m, 7.45 \%/0.1$\mu$m < \textit{Slope A corr} < 13.25 \%/0.1$\mu$m, and the apparent depth between 1.15 and 1.72. 

For comparison, Fig. \ref{SlopesDepths} also shows the centers of regions characteristic for different taxonomic classes based on the measured \textit{Slope A}, \textit{Slope B,} and apparent depths of the mean spectra from the Bus-DeMeo catalogue \citep{2009Icar..202..160D}. The diameter of these regions was estimated from the \textit{Slope A} error based on the one-sigma of the mean S-type spectrum and serves only to improve the  visual interpretation. The region sizes are not variable because not all of the mean spectra in the Bus-DeMeo catalogue have defined characteristic standard deviations.

It was found that the majority of our targets have visible spectra parameters consistent with expected V-type properties. Four asteroids in our sample are dynamically classified as inner-other (IOs): (5524) Lecacheux, (19983) 1990 DW, (51742) 2001 KE$_{55}$, and (90223) 2003 BD$_{13}$. The orbital elements of these bodies are close to the Vesta family (Table \ref{properties}). While all four asteroids show spectral properties associated with the V-type taxonomic class, their visible parameters are quite dispersed, particularly in \textit{Slope A} and the apparent depths (Fig. \ref{SlopesDepths}). 

The spectrum of asteroid (5524) Lecacheux shows slightly higher reflectance in the 0.7 - 0.9 $\mu$m region than expected for a V-type asteroid. The overall shape of the spectrum and the visible parameters are nevertheless fully consistent with V-type taxonomy. Asteroid (5224) presents a low albedo value for a V-type asteroid. However, the albedo determination was based on just four observations using only one of the four filters of WISE (W1). According to \citet{2012ApJ...759L...8M}, in such cases, the derived albedo may be underestimated.

Asteroid (19983) 1990 DW is found with very high \textit{Slope A}, which could indicate a weathered surface for the asteroid. \textit{Slope A} is determined with high uncertainty for this case, due to deviating results when applying different solar analogue stars for the spectrum reduction. Nevertheless, all of the spectral parameters are within an error bar that is consistent with a V-type body, as confirmed by the taxonomic classification using the M4AST tool (Table \ref{slopes}). The visible spectra parameters of asteroid (51742) 2001 KE$_{55}$ are comparable to the control sample region (Fig. \ref{SlopesDepths}) of Vestoids identified by \citet{2016MNRAS.455.2871I}. A surprisingly large apparent depth (1.985 $\pm$ 0.101) found in the spectrum of the asteroid (90223) 2003 BD$_{13}$ could indicate presence of fresh unweathered pyroxene. This is also supported by the low slope in the 0.50 - 0.75 $\mu$m region (\textit{Slope A corr} = 6.73 $\pm$ 0.72).

The six low-i asteroids in our sample, (3188) Jekabsons, (3331) Kvistaberg, (4693) Drummond, (7223) Dolgorukij, (9007) James Bond, and (29733) 1999 BA4, were all identified as V-types. The majority of them, based on their visible spectra parameters (Fig. \ref{SlopesDepths}), fall within the V-type control sample region. For three cases, (3188), (3331) and (29733), we obtained slightly increased apparent depths of their spectra, potentially indicating unweathered surfaces. On the other hand, asteroid (7223) Dolgorukij was found with steep \textit{Slope A} = 20.69 $\pm$ 0.43, suggesting a more weathered surface. 

One asteroid in our sample has been classified as a fugitive, that is, a dynamical group of basaltic asteroids defined by \citet{2008Icar..193...85N} as having $a <$ 2.3 au and $e$ and $i$ comparable to the Vesta family. The asteroid (2275) Cuitlahuac with $a =$ 2.296 au is near the limit of this definition and shows spectral properties similar to most Vesta family asteroids (Fig. \ref{SlopesDepths}). The shape of the spectrum differs slightly from the average V-type spectrum (Fig. \ref{spectra}), with a higher reflectance observed in the 0.75 - 0.90 $\mu$m region. A similar effect can be seen in the spectrum of asteroid (19983) obtained on the same observing night and could be caused by an unsuitable solar analogue star used for the reduction, as deviating results were obtained for different solar analogue stars (an average of the spectra was used for the analysis). Nevertheless, the determined visible spectra parameters are fully consistent with V-type taxonomy.

Overall, our results suggest that in the visible range, low-i, inner-other, and fugitive V-type asteroids do not show recognisable differences when compared to Vesta family bodies. This is in agreement with the results of \citet{2016MNRAS.455.2871I}, who also did not find significant deviations in NIR spectra between the inner main belt dynamical groups. \citet{2020MNRAS.491.5966M} came to a similar conclusion by analysing the distribution of NIR colours of the V-type candidates. It is, therefore, plausible that these bodies originate from (4) Vesta. For fugitives, the dynamical pathways were previously described by \citet{2008Icar..193...85N}. Their simulations, however, did not predict low-i vestoids, meaning that if they originated in Vesta, they were liberated from its surface before or during the Late Heavy Bombardment epoch $\sim$ 3.8 Gy ago \citep{2008Icar..193...85N}.

\section{Visible + NIR spectra analysis of outer main belt asteroids} \label{NIRres}

\begin{figure}
\centerline{\includegraphics[width=.95\columnwidth,angle=0]{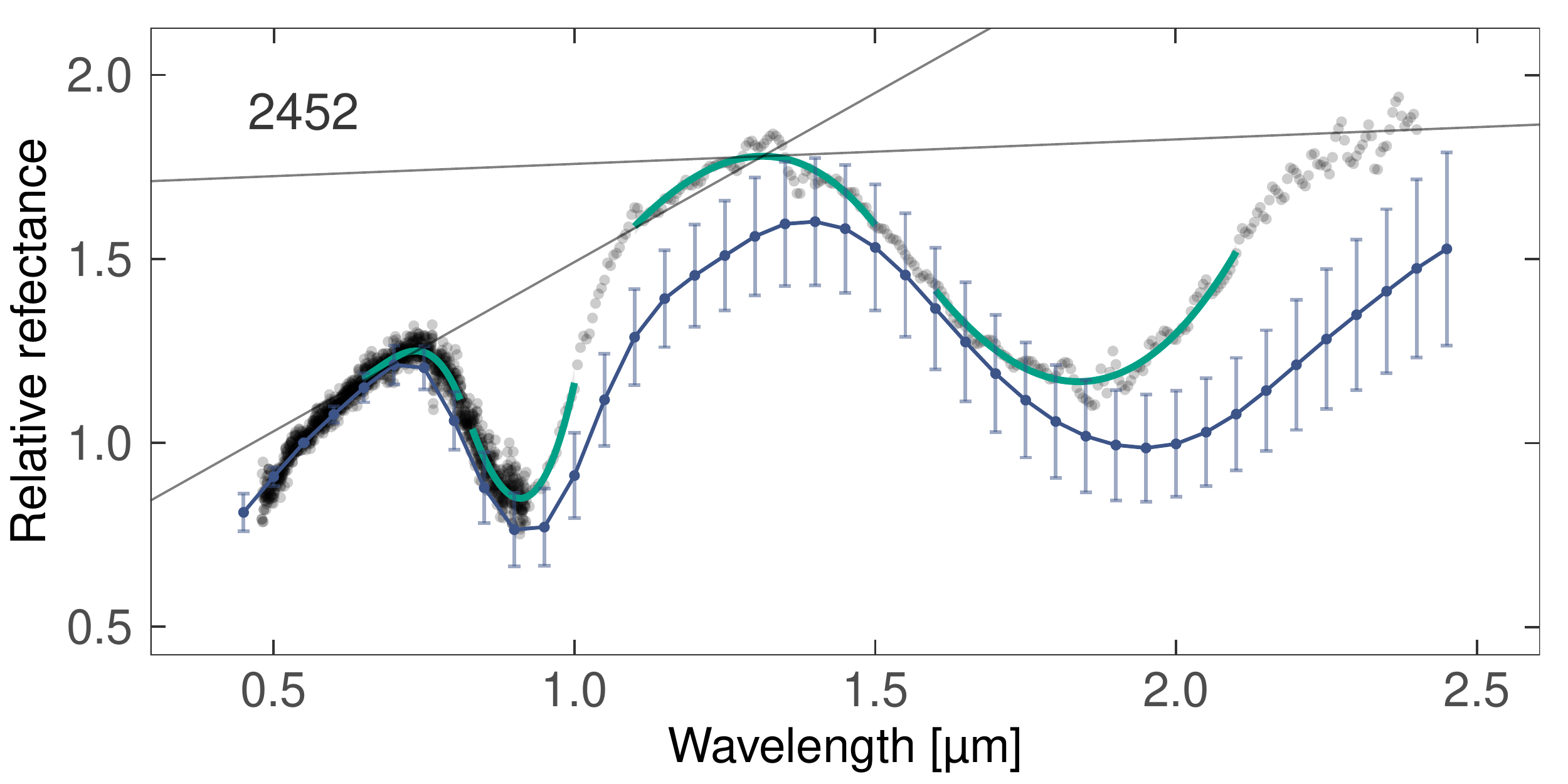}}
\centerline{\includegraphics[width=.95\columnwidth,angle=0]{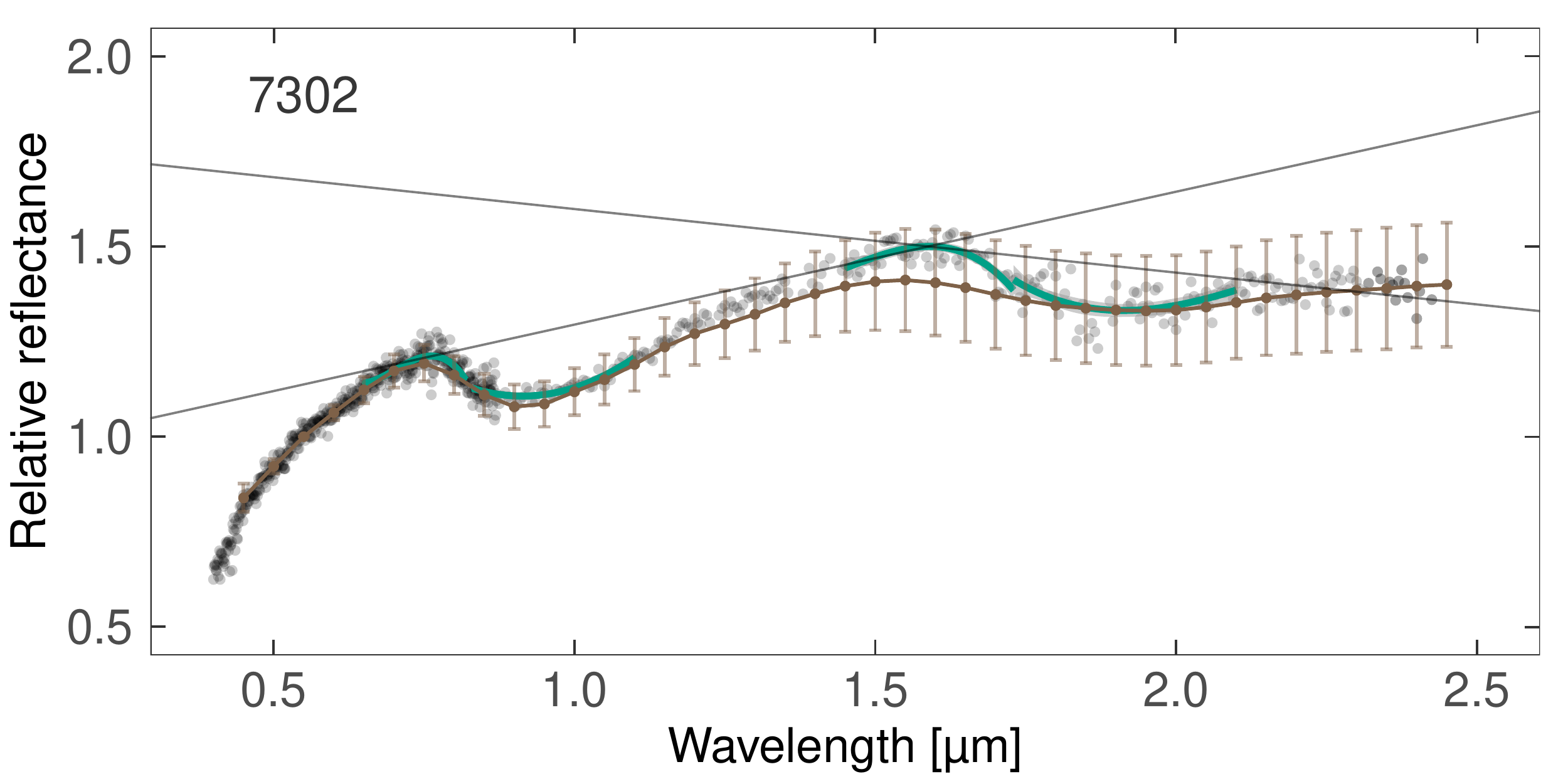}}
\centerline{\includegraphics[width=.95\columnwidth,angle=0]{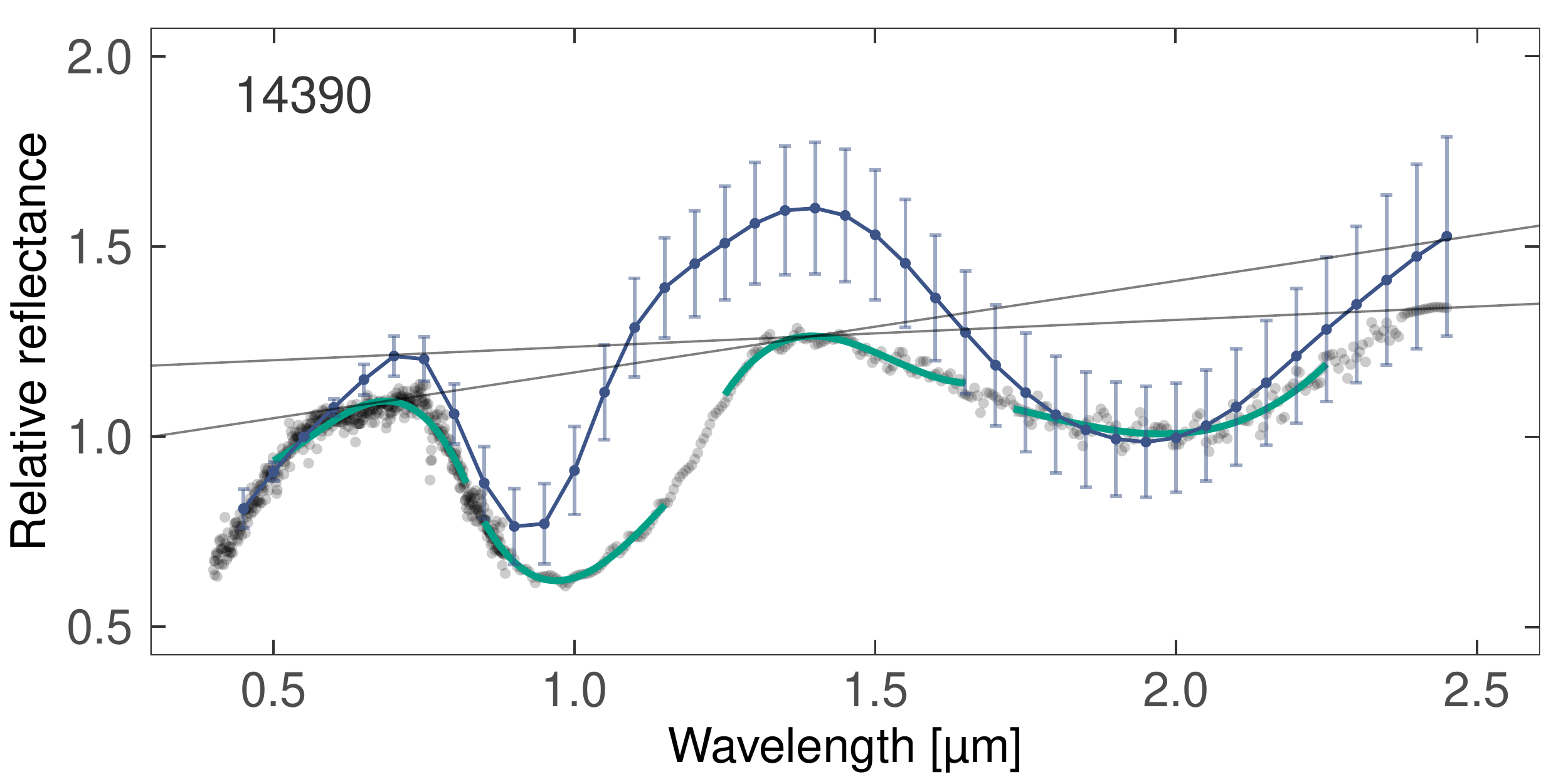}}
\caption[f1]{Visible and near-infrared spectra of the three middle and outer main belt asteroids (MOVs) studied in this work (grey points). The blue and brown curves represent average spectra corresponding to the V-type and S-type taxonomic classes, respectively, and their standard deviation as defined by \citet{2009Icar..202..160D}. The green curves are fourth-order polynomial fits to the maxima and the minima used to compute Slope I and II (straight black lines).} 
\label{VISNIR}
\end{figure}

One of the main interests of this work is the characterisation of V-type asteroids from the middle and outer main belt (MOVs) with no dynamical link to the Vesta family. Therefore, we focus in more detail on asteroids (2452) Lyot, (7302) 1993 CQ, and (14390) 1990 QP$_{10}$, which are all identified as outer main belt asteroids with $a >$ 2.8 au. These asteroids have been identified as atypical cases based on NIR spectra analyses by other authors. Here, we attempt to infer new information about the nature of these bodies based on the acquired visible spectra. In addition to the visible spectra parameters discussed in the previous section, we combined the visible spectra with NIR spectra obtained from the literature (Fig. \ref{VISNIR}) to provide a further compositional analysis based on the determined band centre positions.

\subsection{Parameters}

For the analysis of the composite visible and NIR spectra, we focused on determining the parameters of absorption features near 0.9 $\mu$m and 1.9 $\mu$m caused by the presence of pyroxene minerals, following the methodology first outlined by \citet{1986JGR....9111641C} (also see \citet{2002aste.book..183G} and references therein). These bands (hereafter Band I - BI and Band II - BII) result from the Fe\textsuperscript{2+} electronic transitions in the M1 and M2 crystallographic sites of pyroxene structure \citep{1993macf.book.....B} and their positions are sensitive to both mineral abundance and chemistry \citep{1986JGR....9111641C}. The band centre positions of BI and BII can then be used to determine the molar contents of calcium (Wollastonite) and iron (Ferrosilite), following the equations derived by \citet{2009M&PS...44.1331B}.

First, we computed Slopes I and II defined as straight lines tangent to the relative maxima (black lines in Fig. \ref{VISNIR}) determined from a fourth degree polynomial fit to the spectrum between 0.7 and 1.4--1.7 $\mu$m and between 1.4--1.7 and 2.4--2.5 $\mu$m, respectively (green curves in Fig. \ref{VISNIR}). The last available data point, typically positioned between 2.4 and 2.5 $\mu$m, was used to delimit Band II and compute the Slope II. The slopes were then used to remove the continuum of the spectrum and the wavelength positions of the BI and BII minima were determined as centres of Gaussian fits to the bands using the Fityk software\footnote{https://fityk.nieto.pl/} \citep{Wojdyr:ko5121}. The errors of resulting parameters (Table \ref{NIRparameters}) were determined as standard deviations of individual measurements obtained by adjusting the wavelength range and polynomial degree used to fit the relative maxima.

Finally, we applied a temperature correction to the determined band positions following the procedure from \citet{2009M&PS...44.1331B}. This allows for a direct comparison to properties of HED meteorites measured in laboratory at room temperature. The mean surface temperature of an asteroid at the time of the observation was derived from the energy conservation equation:

\begin{equation}
T = \bigg[\frac{(1-A)L_0}{16 \eta \epsilon \sigma \pi r_{h}^2}\bigg]^{1/4}
,\end{equation}
where $A$ is the bolometric Bond albedo (with mean value of A = 0.15 for V-types from \citet{2014Icar..235...60H} and mean value of A = 0.09 for S-types, following the phase integral values from \citealt{2019A&A...626A..87S}), $L_0$ is the solar luminosity (3.827 x 10$^{26}$ W), $\eta$ is the thermal beaming parameter (assumed to be unity), $\epsilon$ is the infrared emissivity (assumed to be 0.9), $\sigma$ is the Stefan-Boltzmann constant (5.67 x 10$^{-8}$ J s$^{-1}$ m$^{-2}$ K$^{-4}$), and $r_{h}$ is the heliocentric distance of the asteroid at the time of the observation. Then, following \citet{2012Icar..217..153R}, the band centre positions for howardites and eucrites need to be corrected using: 

\begin{equation}
\Delta BI (\mu m) = 0.01656-0.0000552 \,* T (K)
,\end{equation}

\begin{equation}
\Delta BII (\mu m) = 0.05067-0.00017 \,* T (K)
,\end{equation}

while for diogenites, the correction is given as

\begin{dmath}
\Delta BI (\mu m) = 0.0000000017 \,* T^3 (K) - 0.0000012602 \,* T^2 (K) + \\ 0.0002664351 \,* T (K) - 0.0124
\end{dmath}

\begin{equation}
\Delta BII (\mu m) = 0.038544 - 0.000128 \,* T (K)
.\end{equation}

As in other works (e.g. \citealt{2010Icar..208..773M}), we found that the temperature correction for Band I position is negligible. The computed temperatures and corrections for Band II centre position, along with the resulting band parameters for studied asteroids are given in Table \ref{NIRparameters}. Since we found that the taxonomy of asteroid (7302) is closer to an S-type (see discussion in Section \ref{NIRresults}), the correction equations valid for ordinary chondrites \citep{2012Icar..220...36S} were used for the final values, presented in Table \ref{NIRparameters}. 

The methodology introduced by \citet{1986JGR....9111641C} also allows for the determination of the olivine-orthopyroxene abundances based on the ratio of BII and BI areas (BAR). The computed BAR value is also used as an additional check for the basaltic nature of the asteroid, based on its location in the region defined by \citet{1993Icar..106..573G} for a sample of basaltic achondrites in the BAR versus Band I centre space. Considering that the band area ratios are heavily dependent on whether the continuum is subtracted before the computation \citep{2019MNRAS.488.3866M}, we do not provide an analysis of BAR in this work.

\begin{table*}[]
\centering
\small\begin{center}
\caption{Visible and near-infrared spectral parameters based on \citet{1986JGR....9111641C, 1993Icar..106..573G}. The information includes the derived asteroid temperature and the corresponding correction ($\Delta$BII), the Band I and Band II centres (Band II centre includes temperature correction $\Delta$BII) and the slopes.}
\begin{tabular}{ccccccc}
\hline\hline\\[-3mm]
Asteroid& T (K)& $\Delta$BII ($\mu$m)& Band I centre ($\mu$m)& Band II centre ($\mu$m)& Slope I (\%/$\mu$m)& Slope II (\%/$\mu$m) \\
\hline\\[-3mm]
2452  & 147 & 0.029 & 0.915 $\pm$ 0.005 & 1.840 $\pm$ 0.015 & 0.91 $\pm$ 0.07 & 0.06 $\pm$ 0.01  \vspace{0.05cm}\\
7302  & 170 & 0.027 & 0.963 $\pm$ 0.011 & 1.893 $\pm$ 0.018 & 0.31 $\pm$ 0.05 & -0.15 $\pm$ 0.04 \vspace{0.05cm}\\
14390 & 157 & 0.024 & 0.990 $\pm$ 0.012 & 1.977 $\pm$ 0.014 & 0.30 $\pm$ 0.09 & 0.09 $\pm$ 0.04 \vspace{0.05cm}\\
\hline
\end{tabular}
\label{NIRparameters}
\end{center}
\end{table*}

\subsection{Results and discussion}
\label{NIRresults}

\begin{figure}
\centerline{\includegraphics[width=.95\columnwidth,angle=0]{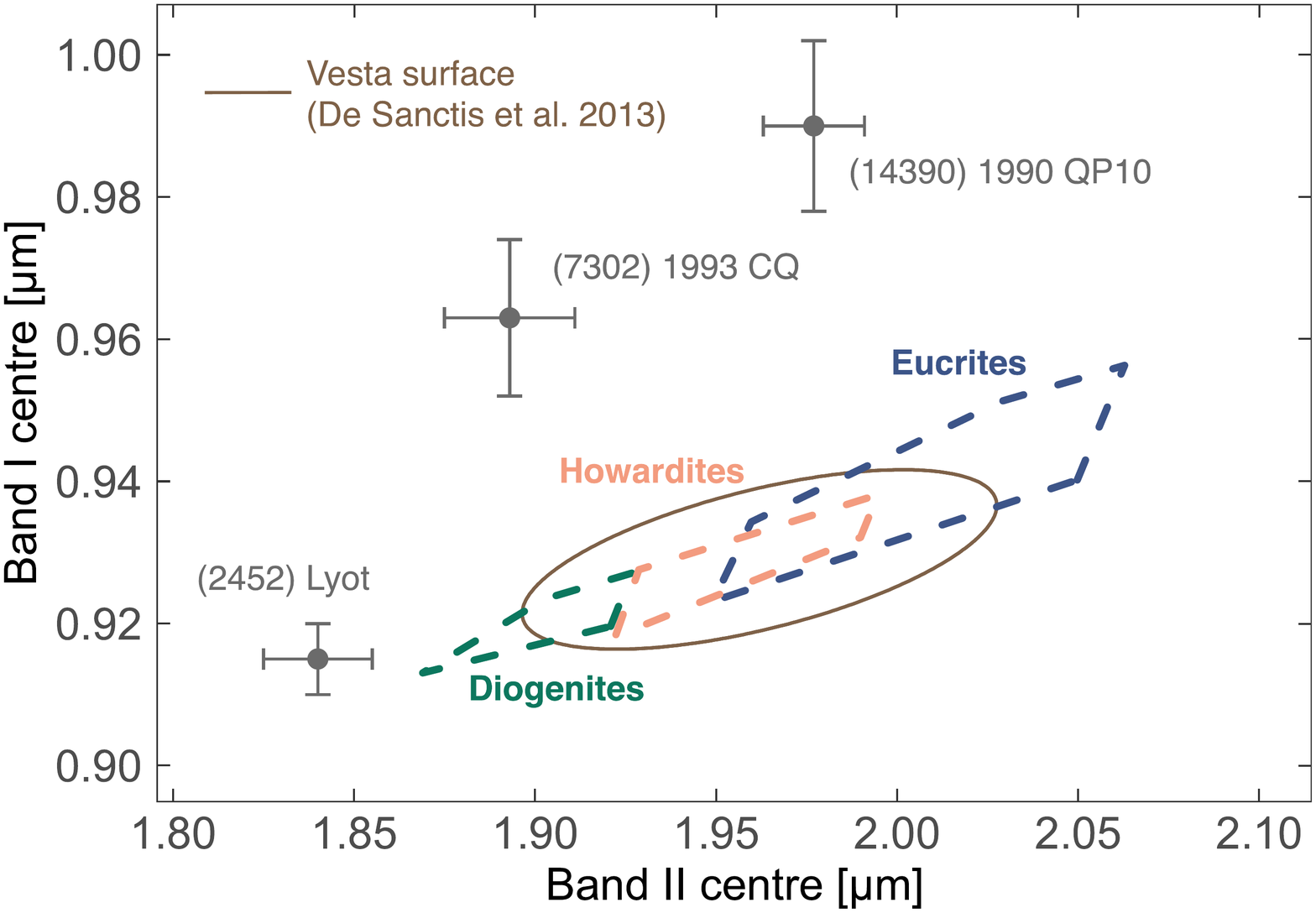}}
\caption[f1]{Band II centre versus Band I centre for the objects studied in this work (grey). Regions outlined by the values computed for Howardites, Eucrites, and Diogenites \citep[from][]{2010Icar..208..773M} are plotted for comparison. The ellipse represents the region occupied by the surface of Vesta as measured by the Dawn mission \citep{2013M&PS...48.2166D}.}
\label{B2vB1}
\end{figure}

Asteroid (2452) Lyot is located in the outer main belt, close to another major confirmed V-type asteroid, namely, (1459) Magnya. Lyot was confirmed as a V-type by \citet{2019MNRAS.488.3866M}, who found that it to be of a diogenite-like composition and significantly different from Vesta and Magnya. The visible spectra analysis confirmed properties consistent with basaltic composition (Table \ref{slopes}, Fig. \ref{SlopesDepths}). Three out of four middle and outer main belt asteroids with visible spectra studied by \citet{2016MNRAS.455.2871I} exhibited increased apparent depths, which could suggest presence of unweathered pyroxene, large grain sizes, or different mineralogies. This is not, however, the case with Lyot, which exhibits an apparent depth of 1.532 $\pm$ 0.047, close to the average V-type values. Furthermore, we combined our visible spectrum with NIR spectrum from \citet{2019MNRAS.488.3866M} and computed band centre positions that are consistent with the results obtained by these authors (Table \ref{NIRparameters}). The obtained Band II position is shifted slightly (1.840 $\mu$m compared to 1.873 $\mu$m), just below the assumed limit for diogenites \citep{2007M&PS...42..235K, 2010Icar..208..773M}. Still, the determined properties are within the error bar consistent with the diogenite-like composition (Fig. \ref{B2vB1}).

The NIR spectrum of asteroid (7302) 1993 CQ was previously studied as a V-type candidate by \citet{2013A&A...552A..85J} and \citet{2018AJ....156...11H}, both finding that the determined band parameters were not consistent with a basaltic composition. The computed slopes and apparent depth from its visible spectrum are indeed not characteristic for V-types but they are, rather, very close to the parameters of an S-type asteroid (Fig. \ref{SlopesDepths}). This is consistent with the results of \citet{2013A&A...552A..85J}, who suggested an ordinary chondrite analogue for (7302). Furthermore, we combined our visible spectrum with the NIR spectrum from \citet{2018AJ....156...11H} to provide a band parameter analysis. The resulting values (Table \ref{NIRparameters}) are consistent with the results of \citet{2013A&A...552A..85J} and \citet{2018AJ....156...11H}, plotted well outside the region associated to V-type asteroids and HEDs in the Band I centre versus Band II centre space, thus ruling out the possibility of a basaltic composition. The overall spectrum fits well with the average S-type spectrum from \citet{2009Icar..202..160D} (Fig. \ref{VISNIR}). It can be therefore concluded that (7302) is an S-type asteroid.

Overall, three NIR spectra have been obtained for asteroid (14390) 1990 QP$_{10}$, all pointing to the conclusion that (14390) is an object difficult to classify and, perhaps, representing a new taxonomic type \citep{2017MNRAS.464.1718M, 2017Icar..295...61L, 2018AJ....156...11H}. Here, we present the first visible spectrum of this body. The spectrum shows a \textit{Slope A} flatter than expected for V-type asteroids (Table \ref{slopes}, Fig. \ref{SlopesDepths}), even more so when corrected for the phase angle effect (\textit{Slope A corr} = 5.44 $\pm$ 1.40 \%/0.1$\mu$m). The very low value for \textit{Slope A} could indicate fresh unweathered surface material. The other visible parameters (\textit{Slope B} = -24.78 $\pm$ 1.97 \%/0.1$\mu$m and apparent depth of 1.685 $\pm$ 0.043) are compatible with V-type asteroids. 

The visible spectrum of (14390) was then combined with the NIR spectrum from \citet{2017Icar..295...61L}, which was chosen as the least noisy from the published NIR spectra. The overall spectrum differs significantly in the NIR region from the expected V-type shape (Fig. \ref{VISNIR}). The spectrum exhibits a very broad band around 1 $\mu$m (Band I) and a shallow band near 2 $\mu$m (Band II), similarly to the results of \citet{2017MNRAS.464.1718M}. The computed Band I centre is at 0.990 $\pm$ 0.012 $\mu$m, which is even higher than the one reported by \citet{2018AJ....156...11H} (0.966 $\pm$ 0.004 $\mu$m) or \citet{2017Icar..295...61L} (0.973 $\pm$ 0.001 $\mu$m). These values are inconsistent with the usual orthopyroxene mineralogy of V-type asteroids \citep{2011MNRAS.412.2318D, 2016MNRAS.455.2871I}, suggesting a substantial abundance of olivine (Fig. \ref{B2vB1}). Furthermore, the spectrum has a very flat Slope I = 0.304 $\pm$ 0.088, which is more similar to an S-type body than to other V-types \citep{2019MNRAS.488.3866M}. Overall, while the V-type is the closest found taxonomic class to fit the visible spectrum of (14390), the spectral differences in NIR are significant and suggest different mineralogy from all other known basaltic asteroids. Therefore, this asteroid remains as unclassified and could represent a new taxonomic class.

Following the work of \citet{2019MNRAS.488.3866M}, our results confirm the high success rate of the MOVIS catalogue in identifying V-type objects. Twelve out of fourteen asteroids studied in this work were confirmed as V-types, yielding an identification success rate of approximately 86\%. As expected, the confirmation of the inner main-belt V-types is more likely due to the presence of the Vesta family and other nearby asteroids which escaped from the Vesta family. In the middle and outer main belt, where the amount of data is more limited, the fraction of false positives increases. Still, as is evident from the results of \citet{2019MNRAS.488.3866M}, a large proportion of the MOVIS candidates from the middle and outer belt includes V-type objects.

\section{Conclusions}  \label{Conclusions}

Based on the analysis of their visible spectra, we confirm 11 new V-type asteroids that are not members of the Vesta collisional family. With respect to their orbital properties, four of these represent inner-other dynamical type: (5524) Lecacheux, (19983) 1990 DW, (51742) 2001 KE$_{55},$ and (90023) 2003 BD$_{13}$. Six V-types have low-i orbits: (3188) Jekabsons, (3331) Kvistaberg, (4693) Drummond, (7223) Dolgorukij, (9007) James Bond, and (29733) 1999 BA4; and one is classified as a fugitive from the Vesta family: (2275) Cuitlahuac. The determined visible spectra parameters suggest presence of fresh surface pyroxene in asteroids (3188), (3331), (29733), and (90223). On the other hand, the spectra of asteroids (7223) and (19983) indicate weathered surfaces. 

In addition, we analysed three peculiar outer main belt V-type candidates based on their visible + NIR spectra. We confirm the previously reported diogenite-like composition of (2452) Lyot, which differs from the bulk composition of Vesta and Magnya. The spectrum of asteroid (7302) is inconsistent with a basaltic composition, as previously suggested from NIR analyses. We report that (7302) is likely an S-type body. 

We provide the first visible spectrum of asteroid (14390) 1990 QP$_{10}$. Three previous NIR studies reported this asteroid as an unusual object of an unclassified taxonomic type. The visible spectrum has apparent depth and slope in the 0.80 -- 0.92 $\mu$m region, similar to the properties of V-type asteroids, but it has flatter slope in the 0.50 -- 0.75 $\mu$m region. The overall visible + NIR spectrum is significantly different from V-type taxonomic type, with a very broad band near 1 $\mu$m and a shallow band near 2 $\mu$m. The spectrum of asteroid (14390) suggests a different mineralogy from other known basaltic asteroids, potentially representing, therefore, a new taxonomic type.

Overall, our results demonstrate the efficiency of the MOVIS catalogue in identifying V-type objects, with a success rate of over 85\%. The identification of V-types in the inner main-belt is more likely due to the presence of the Vesta family and other nearby asteroids which escaped from the family. The amount of data is more limited in the middle and outer main belt, leading to an increase in the fraction of false positives. Earlier results from \citet{2019MNRAS.488.3866M} have, however, shown that a large fraction of the MOVIS candidates from the middle and outer belt include V-type objects.

\begin{acknowledgements}

PM was supported by the ERASMUS+ project 2017-1-CZ01-KA203-035562, the Slovak Research and Development Agency grant APVV-16-0148 and the Slovak Grant Agency for Science grant VEGA 01/0596/18. JdL acknowledges financial support from the project ProID2017010112 under the Operational Programmes of the European Regional Development Fund and the European Social Fund of the Canary Islands (OP-ERDF-ESF), as well as the Canarian Agency for Research, Innovation and Information Society (ACIISI), and the project AYA2017-89090-P of the Spanish MINECO. HM acknowledges financial support of the fellowship 300954/2020-4 from National Council of Scientific and Technological Development (CNPq) - Brazil. J-AM acknowledges the support of the Romanian National Authority for Scientific Research - UEFISCDI, project number PN-III-P1-1.2-PCCDI-2017-0371.
     
\end{acknowledgements}

\bibliographystyle{aa}
\bibliography{references}

\end{document}